RESEARCH ARTICLE

# Dynamic Advisor-Based Ensemble (dynABE): Case study in stock trend prediction of critical metal companies

Zhengyang Dong *

Middlesex School, Concord, Massachusetts, United States of America

* zdong@mxschool.edu

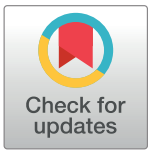

## Abstract

Stock trend prediction is a challenging task due to the market's noise, and machine learning techniques have recently been successful in coping with this challenge. In this research, we create a novel framework for stock prediction, Dynamic Advisor-Based Ensemble (dynABE). dynABE explores domain-specific areas based on the companies of interest, diversifies the feature set by creating different "advisors" that each handles a different area, follows an effective model ensemble procedure for each advisor, and combines the advisors together in a second-level ensemble through an online update strategy we developed. dynABE is able to adapt to price pattern changes of the market during the active trading period robustly, without needing to retrain the entire model. We test dynABE on three cobalt-related companies, and it achieves the best-case misclassification error of 31.12% and an annualized absolute return of 359.55% with zero maximum drawdown. dynABE also consistently outperforms the baseline models of support vector machine, neural network, and random forest in all case studies.

## 1 Introduction

Stock trend prediction is an area of interest to researchers and investors alike, due to the complex patterns underlying the price data and the high profitability of successful trading strategies. In recent years, machine learning has become a popular technique for modeling the stock market. There are three quantitative approaches to stock prediction in general, each exploring different areas related to the stock market. The most common approach is based on general indicators, specifically the historical price and technical indicators. Such an approach relies on the traditional chartist theory that price patterns in the past will reoccur in the future [1]. The second approach is based on sentiment analysis, using natural language processing techniques to interpret text-based data like news articles. It is based on financial research such as He *et al.* [2] who show how investor sentiment influences stock returns. The third approach is based on the intercorrelation of corporations that uses information of other companies to predict the stock trend of one company, such as the recent work by Chen and Wei in 2018 [3].

In this paper, we present a novel stock prediction model, Dynamic Advisor-Based Ensemble (dynABE). There are four main contributions of our research to current works: the The symbols of the specific data in the database used in this research are provided as supplementary files and described in the paper.

**Funding:** The author received no specific funding for this work.

**Competing interests:** The author has declared that no competing interests exist.

exploration of domain-specific information for high-frequency predictions; the establishment of an effective, first-level ensemble learning framework; the proposal of "advisors" for a second-level ensemble; and an online update strategy for dynamic flexibility.

First of all, instead of the three common approaches to stock prediction, this work is one of the few that explores the direction of commodity-stock relationship by incorporating *domain-specific information*. By "domain-specific information," we mean the information that is related to the specific industry of a certain company. For example, the automobile market would be a type of domain-specific information for automobile producers, clean energy technologies for oil mining companies, and the consumer electronics market for technology companies. While fundamental analysis often explores the industry of specific companies to estimate their intrinsic values for *long-term* investments [4], few works use quantitative and high-frequency domain-specific information for *short-term* stock prediction.

Moreover, as we will later show in the literature review, no single machine learning model has been established to be superior for stock prediction. This calls for the need of ensemble learning, which combines the strengths of different models to compensate for one another's mistakes when no single model is guaranteed to be most effective [5]. Our work presents an effective model ensemble framework that has a hybrid feature selection method and uses stacking to combine the base models.

In addition, we propose the concept of "advisors," which is especially effective for the stock market. Specifically, we first define a number of domain-specific areas we want to investigate for a certain company. For each area, we find a pool of related features that will go through the previously defined ensemble learning framework to form one advisor. The multiple advisors are then combined to form a second-level ensemble.

The last innovation lies in the method we designed for combining these advisors, an online update strategy performed during the active trading period. Most current methods for stock prediction are static after the initial training. Therefore, they lack the flexibility to update themselves during the active trading period, making them vulnerable to the stock market dynamics—the market's changes in price patterns may render a previously effective prediction model suddenly less accurate. In contrast, dynABE uses an online update strategy to dynamically weigh the advisors during trading. We will later elaborate the details of the online update. Intuitively, the use of advisors and online update ensures that all factors of the stock market that we wish to investigate, as defined in the formation of advisors, are available at all times during trading. Therefore, even if the price pattern of the stock market changes, dynABE is still able to adapt to the new pattern by changing the weights of the advisors. We show in our experiments that this additional dynamic flexibility of dynABE effectively increases its accuracy. In addition, since we do not need to retrain the base models with new data but only update the weights of the advisors, dynABE's online update method is robust with few parameters.

We compare the performance of dynABE to three baseline models commonly used for stock prediction, namely support vector machine, neural network, and random forest. We show that dynABE consistently outperforms all the baseline models in all our case studies. We further use the predicted stock trends as trading signals on a naïve trading strategy to illustrate dynABE's high potential profitability. It is important to point out that while the domain-specific information we use does have more financial motivations, it does not nullify the effectiveness of general indicators, because financial motivation does not guarantee predictive superiority. In addition, dynABE works well with domain-specific data because it is easier to group the features into advisors using financial knowledge. However, one can certainly use dynABE with other data as long as the feature set can be effectively grouped into advisors. In other words, dynABE is not limited to domain-specific information, and we will discuss its generalization in this paper as well.

## 2 Related works

We examine the existing literature based on the different machine learning models they use. Early works often used simple regression methods to estimate stock returns. For example, Fama *et al.* [6] used a least square regression to predict the NYSE stock portfolio stock returns, and Pesaran *et al.* [7] predicted the excess returns for S&P 500 and Dow Jones Industrial portfolio stocks using multivariate regression. More complex models such as support vector machine soon became popular, such as Lee's work [8] that applied a hybrid feature selection method on general indicators and fed the selected features to a support vector machine for prediction of the NASDAQ index. Support vector machine is also commonly used for sentiment analysis, such as Schumaker and Chen's work [9] and Hagenau *et al.*'s work [10] that both used support vector machine to interpret news articles.

Neural network is another popular model that explores the nonlinearity of stock data. As early as 1998, Saad *et al.* [11] have already compared time delay, recurrent, and probabilistic neural networks for stock prediction. While both Tsang *et al.* [12] and Tsai *et al.* [13] have had successes using a basic three-layer vanilla network structure, more recent works use neural network variants that are sometimes termed as "deep learning models." Nelson *et al.* [14] in 2017 and Das *et al.* [15] in 2018 have both used long short-term (LSTM) recurrent neural networks, respectively on general indicators and sentiment analysis. In addition, Ding *et al.* in 2015 [16] and Chen and Wei in 2018 [3], the most recent work in our literature review, have both used convolutional neural networks, respectively on sentiment analysis and intercorrelation of corporations.

Despite the successes of support vector machine and neural network, they are still sometimes outperformed by other methods in comparative studies. For example, Patel *et al.*'s research [17] compared performances of neural network, support vector machine, random forest, and naïve-Bayes classifier for stock trend prediction, and random forest was the most accurate. Similarly, in the comparative study of Ballings *et al.* [18], random forest outperformed models including logistic regression, neural networks, k-nearest neighbor, and support vector machine as well as other ensemble methods including AdaBoost and Kernel Factory. Therefore, as previously mentioned, the very diversity of successful models for stock prediction suggests that there is no single well-established model proven to be most effective. Table 1 summarizes the related works. Besides the machine learning methods, we have also listed the general approach of each work, which is among general indicators, sentiment analysis, and intercorrelation of corporations.

## 3 Data selection

### 3.1 Critical metal companies as the case study

We choose critical metal companies as the case study to test dynABE, specifically cobalt companies, due to the growing investment interests in critical metals [21]. In recent years, the demand for metals in technology has been shifting from several major metals, such as iron and copper, to numerous minor metals, such as cobalt and indium. Many of these minor metals do not constitute their own ores but exist in low concentrations in ores of other common metals. Therefore, their high susceptibility to supply instability and industrial importance lead to the term "critical metals" [22]. Cobalt is a main critical metal for its rising importance in technology. According to the United States Geological Survey (USGS), in 2017, the largest use of cobalt was for superalloys in aircraft engines in the United States and the rechargeable battery

Table 1. Summary of related works. Works are grouped by the machine learning models they utilize.

| Author (s) | General Approach | Dataset | | Machine Learning Model |
|---|---|---|---|---|
| | | Features | Target | |
| Fama et al. (1988) [6] | General Indicators | Dividend yield | NYSE portfolio stock returns | Linear regression |
| Pesaran et al. (1994) [7] | General Indicators | Dividend yield, interest rates, inflation rates, and industrial production index | S&P 500 and Dow Jones Industrial portfolio stock returns | Linear regression |
| M.-C. Lee (2009) [8] | General Indicators | Future contracts, spot indices, and previous day's NASDAQ index | NASDAQ index | Support vector machine |
| Schumaker and Chen (2009) [9] | Sentiment Analysis | Financial news articles | S&P 500 companies | Support vector machine |
| Hagenau et al. (2013) [10] | Sentiment Analysis | Corporate announcement news | Selected companies | Support vector machine |
| Saad et al. (1998) [11] | General Indicators | Historical prices | Selected companies | Time delay, recurrent, and probabilistic neural networks |
| Tsang et al. (2007) [12] | General Indicators | Historical prices | HSBC stock trend | 3-layer neural network |
| Tsai et al. (2010) [13] | General Indicators | Financial and macroeconomic indices | TSE listed companies | 3-layer neural network |
| Ding et al. (2015) [16] | Sentiment Analysis | News events | S&P 500 index and individual companies | Convolutional Neural Network |
| Nelson et al. (2017) [14] | General Indicators | Hitorical prices and technical indicators | Companies listed in IBovespa index | Recurrent (LSTM) Neural Netwok |
| Das et al. (2018) [15] | Sentiment Analysis | Twitter and other streaming data | Google, Microsoft, and Apple | Recurrent (LSTM) Neural Network |
| Chen and Wei (2018) [3] | Intercorrelation of Corporations | Corporate information | 2988 companies listed in the "tushare" API | Convolutional Neural Network |
| Patel et al. (2015) [19] | General Indicators | Technical indicators | Two stock prices and two stock indices | Comparison between ANN, SVM, random forest, and naïve-Bayes |
| Ballings et al. (2015) [20] | General Indicators | Financial indices, corporate information, and economic indicators | Stock trend of 5767 listed European companies | Comparison between logistic regression, neural networks, k-nearest neighbor, SVM, random forest, and AdaBoost |

https://doi.org/10.1371/journal.pone.0212487.t001

industry in China [23]. Indeed, according to a recent study by Olivetti et al. [24] in 2017, many of these high technology industries are materials dependent because they are directly enabled by the availability of certain materials. It is not surprising that, as these high technologies continue to develop, the Cobalt Development Institute forecasts a 68% increase in cobalt consumption from 2015 to 2025 [25]. Moreover, cobalt's average annual price in 2017 also doubled due to such "strong demand from consumers, limited availability of cobalt on the spot market, and an increase in metal purchases by investors," according to the report of USGS [23].

We choose three cobalt-related companies to represent cobalt miners, cobalt refiners, and miners of cobalt's carrier metals. The three companies are Jinchuan Group International Resources (HKG: 2362) listed in Hong Kong Stock Exchange (HKSE), Sumitomo Metal Mining (TYO: 5713) listed in Tokyo Stock Exchange (TSE), and Zijin Mining Group (HKG: 2899) listed in HKSE. Jinchuan Group represents mining companies that have cobalt as their main metal product. As a leading global cobalt miner, Jinchuan has an annual mining capacity of 10,000 tons that constitutes around 10% of the world production [26]. Sumitomo Metal Mining represents companies that refine cobalt as part of their business. Sumimoto Metal Mining focuses on nickel and cobalt refining and is the only electrolytic cobalt refiner in Japan [27], having produced 4,000 tons of electrolytic cobalt in 2017 [28]. Zijin Mining Group represents companies that do not directly produce cobalt but have carrier metals of cobalt as their main metal products. In this case, Zijin Mining is the second largest mined copper producer in China [29], and cobalt is often produced as a byproduct of copper [22].

For this research, historical data from the entire year of 2015 is used for training. Data of 2016 and the first half of 2017 is used as the validation set, which the model has not previously seen. Therefore, the validation set is the simulated active trading period of the model. We test the model's prediction accuracy during validation and the profitability of a naïve trading strategy using the prediction results as trading signals.

### 3.2 Feature set makeup and advisors

One key innovation of dynABE is the concept of advisors. We first use financial knowledge to determine general areas that we think will influence a certain company's stock price. Then we find a pool of features for each area, which will be candidate features for one advisor. Therefore, the advisor is responsible for this one area. The term "advisor" is an analogy that when executives make decisions, they never try to understand the implications of every single possible factor themselves. Instead, they listen to suggestions of different advisors, each an expert in his or her field, in order to consider a variety of factors. Similarly, in our model, each advisor only investigates features in one particular area. The "executive decision" is made in the final step when all the advisors are combined to generate the final prediction. Since we investigate domain-specific information in this research, we create three advisors to reflect the commodity-stock relationship for critical metal companies in three areas: the commodity (metal) market, cost of production, and macroeconomics.

The first advisor investigates the metal market. The price data of the direct metal products of the critical metal companies is most apparently relevant. This is because, despite the firms' hedging, the stock prices of mining companies still have a degree of price exposure to their product metals [30]. Therefore, the prices of the product metals influence the revenues of the mining companies and thus their intrinsic values, so their stock prices are affected. Moreover, we also include information about metals that are not direct metal products of those companies, because the metal market is itself intercorrelated. For example, Fu *et al.*'s research in 2018 [31] shows that there is a strong supply constraint for carrier metals on their byproduct metals, so prices of different metals can be related to each other.

The second advisor reflects the cost of production for mining. We incorporate commodity prices of chemicals used in industrial productions and prices of energy resources. Energy is especially important to the mining industry as crude oil and electricity are considered to be the main operating costs of mining [32]. Therefore, these raw material commodities for mining affect the profits of the metal producers. In addition, crude oil has price spillovers with many markets, including the stock market, and Xiao *et al.* [33] have shown that oil price uncertainty influences stock returns.

Last but not least, the third advisor includes macroeconomic factors, often shown to influence the entire stock market [34]. As we explore the commodity-stock relationship, we use various commodity indices. Most commodity indices we include, such as the London Metal Exchange Index (LMEX) [35], are average commodity prices to reflect a larger commodity market of interest like the metal market. Therefore, these are a macroscopic way to examine relevant commodities to metal companies' stock prices. In addition, we consider currency exchange rates as an important macroeconomic factor. We mostly include exchange rates between major metal producing countries and the United States, which we take as a major metal consumer. Therefore, the currency exchange rates determine the actual value of the profit in the producer's country when the producer sells to a foreign country. Metal prices and currency exchange rates of major metal-producing countries are also sometimes correlated. As an example, it was found that Chilean Peso to US Dollars exchange rate and copper prices are highly correlated, with a correlation coefficient of 0.93 from July 7 to September 7 of 2015 [36].

For this research, we have obtained all the above-mentioned data from the Refinitiv Datastream [37] service, which is only available with subscription. However, we have provided the description of each feature in the feature set and its corresponding symbol in the Refinitiv Datastream database in S1 File of supporting information. Therefore, one can replicate the experiments by gaining access to the Refinitiv Datastream and using the symbols provided. Alternatively, one can also use a different service to obtain a similar feature set.

### 3.3 Proposed data selection standards

The previous section illustrates how we choose the advisors and the feature set for the critical metal case study. Here we propose a general standard for data selection if one wishes to implement dynABE. In order to make a financially sound dynABE model, one needs to identify a number of factors that influence the stock price of the companies of interest. For example, we chose three commodity-related factors, the commodity market, raw material commodities (cost of production), and macroeconomics, because this research focuses on the commodity-stock relationship. Then one can select a pool of features for each factor to include as many features in each pool as available: there is no need for the manual filtering of features because we entrust the future feature selection to a computational procedure introduced later. Therefore, the real data selection task lies in identifying these factors. There is indeed a high degree of freedom in choosing factors one wishes to investigate, but since the purpose of this research is to demonstrate that the domain-specific direction of the commodity-stock relationship is effective, experimenting with other factors is beyond the scope of this paper. Therefore, we recommend creating advisors based on the same commodity-stock relationship to this paper, i.e. three advisors respectively for the commodity market of the new company of interest, its raw material commodities (cost of production), and macroeconomics.

Alternatively, the advisor creation does not always have to be financially sound. We hypothesize that if dynABE were to be used for time-series prediction in general, one can create the advisors purely computationally. For example, one can use some clustering algorithm on a very large feature set and group one advisor for each cluster of features. However, this guess requires experimental validations in the future.

## 4 Methodology

Here we give a detailed description of dynABE after the advisors have been chosen. There are two main parts of dynABE, the ensemble learning framework for each advisor and the online update strategy for combining the advisors. We first introduce our ensemble learning framework, including feature selection, base models, ensembles of the base models, and the optimization of bootstrap aggregation. Then we discuss the online update strategy and provide the algorithms for implementation. We do not elaborate the data cleaning and preprocessing steps, though we include them as S1 Appendix in supporting information. Fig 1 is an overview of dynABE, and Fig 2 illustrates the ensemble learning framework for an individual advisor.

Note that there are two places where we combine different models, the ensemble of base models and the online update of advisors, yet the two serve different purposes. Ensemble is combination at the algorithmic level for joining the strengths of different machine learning models. On the other hand, online update combines observations from different fields of economics, and it helps the model from a financial perspective. Both are necessary, non-redundant parts of dynABE.

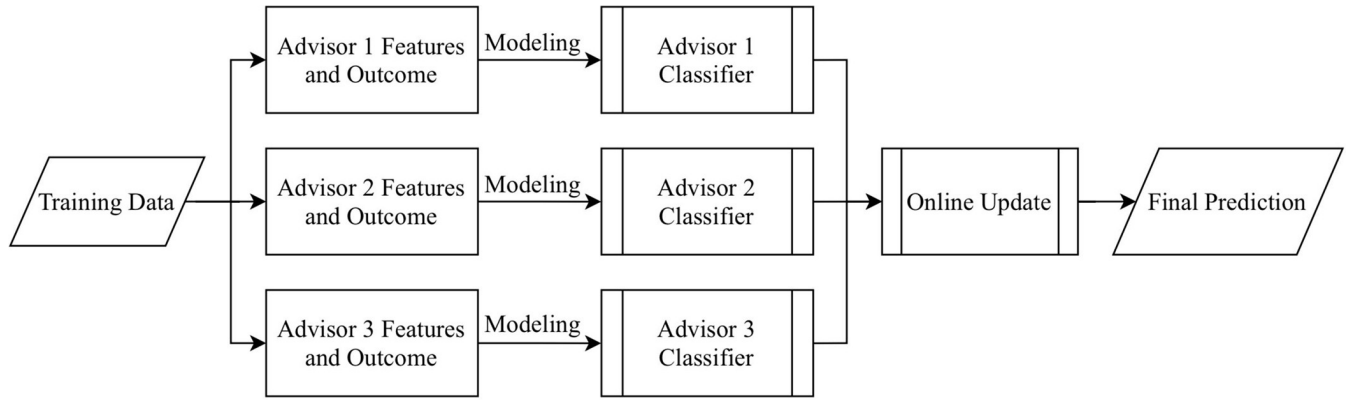

**Fig 1. Overview of dynABE.**

https://doi.org/10.1371/journal.pone.0212487.g001

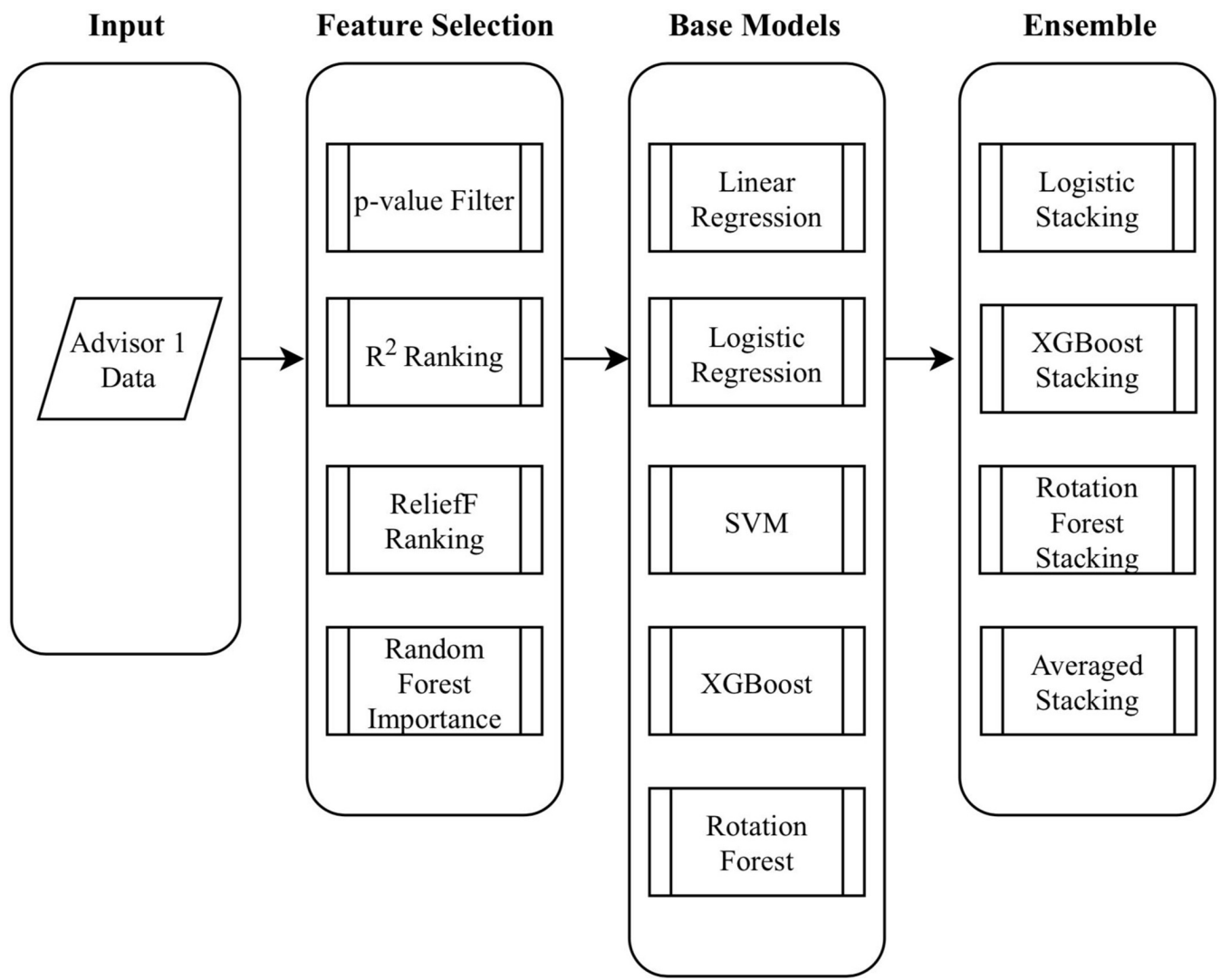

**Fig 2. dynABE's ensemble learning framework for one advisor.**

https://doi.org/10.1371/journal.pone.0212487.g002

### 4.1 Feature selection

Since we start with a large number of features, dimensionality reduction is necessary. Many works have shown that machine learning models work better with lower dimensional but more representative data, such as Huang *et al.*'s works [38] [39] that improve the accuracy of financial early warning models through a kernel entropy manifold learning algorithm. For our work, we perform dimensionality reduction with a feature selection process we designed. The feature selection step is also an ensemble of individual feature selection methods, which is not uncommon in literature [40]. Our strategy is to combine the feature rankings of different feature selection methods. All features first undergo a p-value filter to remove statistically insignificant features. Then the filtered feature set goes through $R^2$ ranking, RELIEF ranking, random forest importance ranking, and the rankings are finally combined together.

As a rough filter, we perform a univariate linear regression between each feature and the outcome of the stock trend in the training period. Then, after obtaining the p-value of each feature, we filter out those with p-values greater than 0.5. While by custom, the choice of p-value thresholds is often smaller, such as 0.01 or 0.05 [41], we want to leave more space for future steps. Therefore, we choose a bigger p-value threshold to make this filter more tolerant.

After filtering, the first feature ranking method is $R^2$ ranking. $R^2$ can be interpreted as the proportion of the variance of output observations that can be predicted from input observations [41]. We extract the $R^2$ value of each univariate regression model fitted between each feature and the outcome and rank the features accordingly.

The second feature ranking method is ReliefF [42], a heuristic measure of features which is an extension of the RELIEF measure proposed by Kononenko [43]. RELIEF favors features whose values distinguish the most in similar observations of different classes [42]. ReliefF optimizes RELIEF by changing certain measures but adopts the same general idea.

The third feature ranking method is random forest importance, a variable importance measure used in random forest. Like most tree-based methods, random forest assigns a variable importance measure to each feature. We use the mean decrease in accuracy, which calculates the decrease in accuracy when the value of a feature is permuted. Thus, the more significant a feature is, the more decrease in accuracy we should observe.

Finally, we combine the results of these three rankings with majority vote. We adopt a simple voting strategy: for any feature, we average its ranks in the three ranking methods to obtain its combined rank, and whichever feature with the highest combined rank would be ranked as the most important feature. We keep the top 20% as the selected feature set. The selected features are included in S2 File in supporting information for experimental replications.

### 4.2 Base models and ensemble

With the selected feature set, we first build five base models and then combine them using different ensemble methods. Unless the original paper is cited, the explanations of the models in this section are based on Hastie, Tibshirani, and Friedman's book, *Elements of Statistical Learning* [44]. Because most base models are common in machine learning, we only briefly introduce them. We also discuss the hyperparameter tuning process for each model.

**4.2.1 Linear regression.** Linear regression with elastic net regularization is used as the first base model. Elastic net regularization is a weighted combination of L1 and L2 regularization. The objective function is: $argmin_{\boldsymbol{\beta}}\{\|\boldsymbol{y} - \boldsymbol{X}\boldsymbol{\beta}\|^2 + \lambda(\alpha\|\boldsymbol{\beta}\|^2 + (1-\alpha)\|\boldsymbol{\beta}\|_1)\}$, where $\lambda$, the regularization parameter, is greater than 0, and $\alpha$, the relative strength of L2 regularization compared to L1, is between 0 and 1. We first grid search possible values of $\alpha$ from 0 to 1, incrementing by 0.1 at each step. With a fixed $\alpha$ at each step, we then tune $\lambda$ with a 10-fold cross-validation using mean squared error to obtain the optimal $\lambda$ for this certain $\alpha$.

**4.2.2 Logistic regression.** Logistic regression is the second base model, which is an extension of linear regression fitted on logistic functions so that the model outputs the probability for classification problems. Because logistic regression has the same hyperparameters as linear regression, $\lambda$ and $\alpha$ are tuned through grid search in the same way.

Both linear regression and logistic regression are simple machine learning models compared to recent ones in the field. We incorporate them as base models nonetheless for two reasons. First, the assumption of ensemble learning is that the combination of weak classifiers can boost the prediction performance, so regardless of how weak linear and logistic regressions seem, as long as they are nontrivial and perform better than random guessing, they are worth to be incorporated as base models. In addition, because linear and logistic regressions are simpler than other complex base models, they are likely to make very different predictions, which adds to the diversity of the ensemble.

**4.2.3 Support vector machine.** Support vector machine (SVM) is the third base model. It separates samples of different classes with a hyperplane and maximizes the margin each class is from the plane. It is formulated as a constraint optimization problem and solved with a Lagrange multiplier. The solution of the Lagrange multiplier only depends on the dot products of pairs of sample points ($x_i$ and $x_j$). SVM then replaces the dot product of $x_i$ and $x_j$ with the inner product $\langle h(x_i), h(x_j) \rangle$ in the Lagrangian, where $h(x_i)$ and $h(x_j)$ are non-linear transformations of $x_i$ and $x_j$. In this way, SVM expands the feature space into higher dimensions.

We choose the radial basis kernel, which defines $\langle h(x_i), h(x_j) \rangle = e^{-\gamma |x_i - x_j|}$. In addition to $\gamma$, the coefficient in the radial basis function, we also tune $C$, which defines how much tolerance we have towards the case when the two classes cannot be perfectly separable by a hyperplane in the given dimension. We tune these two hyperparameters using grid search.

**4.2.4 Extreme gradient boosting (XGBoost).** XGBoost is the fourth base model, which uses a classification and regression tree (CART) ensemble based on gradient boosting developed by Chen *et al.* [45]. XGBoost is trained additively. At each step, a new CART model is added to the ensemble to improve the model performance. Gradient boosting greedily reduces its error by setting the negative gradient of the ensemble's error to be the objective of the next CART being added.

There is a variety of important hyperparameters to tune for XGBoost. We specifically focus on 8 hyperparameters: the learning rate ("eta"), minimum sum of instance weights ("min_child_weight"), maximum depth of a tree ("max_depth"), row subsample rate ("subsample"), column subsample rate ("colsample_bytree"), L1 and L2 regularization terms on weights ("alpha" and "lambda"), and minimum loss reduction ("gamma"). Details of these hyperparameters can be found in the original paper. Because there are so many hyperparameters, grid searching through all possible combinations is not realistic. Therefore, we adopt the random search strategy [46] for tuning, which is especially useful when the search space is very big.

**4.2.5 Rotation forest.** Rotation forest, another CART ensemble method, is the final base model. Rotation forest was proposed by Rodríguez *et al.* [47] to complement existing ensemble strategies. According to the authors, common ensemble methods either increase base model accuracy while sacrificing base model diversity, such as boosting methods, or increase base model diversity while sacrificing base model accuracy, such as random forest [47]. Rotation forest is claimed to preserve base model accuracy and diversity at the same time.

In rotation forest, the complete feature set is randomly split into $K$ subsets. Each subset undergoes a principal component analysis (PCA), and all principal components are kept as the new features for this subset, hence the "rotation" of features. The rotated features from each subset form one CART $D_1$. This process is repeated $L$ times, each time training a different CART $D_i$. The final prediction is averaged over all CARTs $D_1, \ldots, D_L$. The diversity of rotation

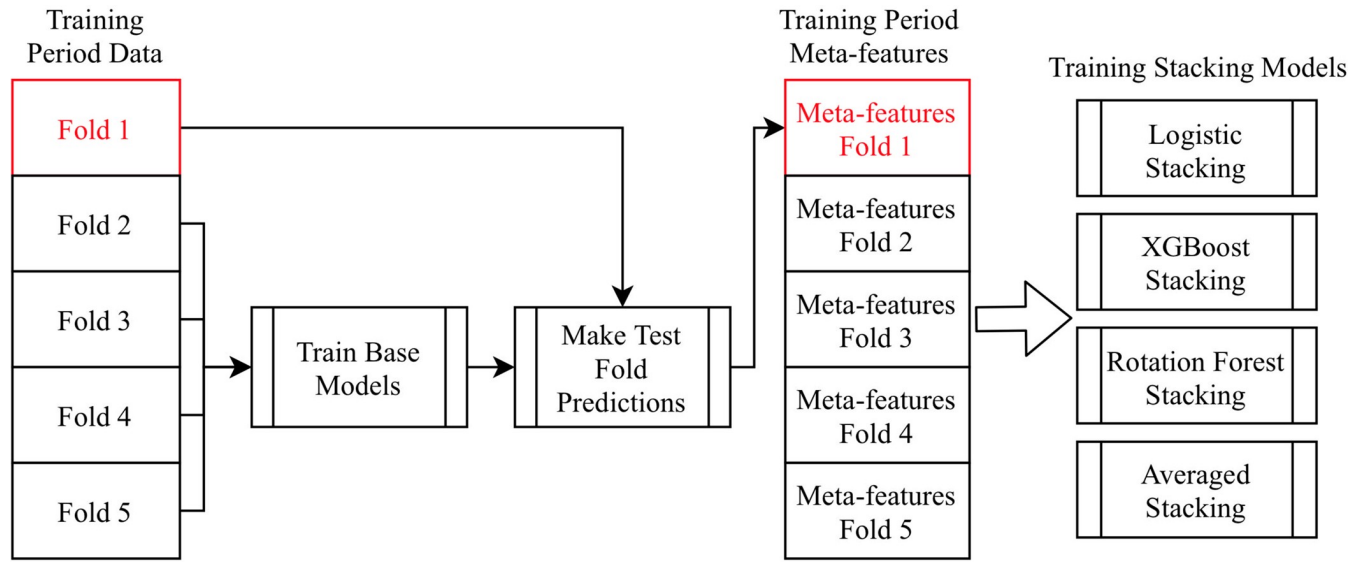

**Fig 3. The training process of stacking.** We only show training the first-fold as an example, which is highlighted in red.

https://doi.org/10.1371/journal.pone.0212487.g003

forest is achieved through the random splits in creating feature subsets that generate different rotations. And since rotation forest keeps all the principal components, the model accuracy is not sacrificed.

Rotation forest only has two hyperparameters, the number of CARTs in the ensemble and the size of the feature subsets for splitting. In practice, we find that hyperparameter tuning of rotation forest does not affect its performance a lot, so we fix the number of classifiers to 10 and the size of each feature subset to be 3.

**4.2.6 Base model ensembles with stacking.** After the base models are trained, we put them into an ensemble using the stacking method. Stacking puts the predictions of the base models as new features into another model to discover relationships between each base model prediction. It is crucial to avoid using the entire training data to train both the base models and the stacking models, which leads to overfitting because the same data is used twice. Instead, we perform a 5-fold cross-validation where all base models are re-trained at each fold to make a cross-validated training set prediction. This would be the meta-features for the training period. Since the base models have not seen the validation data, we directly use the base model validation predictions as meta-features for the validation set.

We use XGBoost, logistic regression, and rotation forest as the stacking models, and each stacking model makes a prediction on the validation set. We further add another validation prediction by averaging the three previous predictions and call it averaged stacking. We make each advisor generate multiple stacking predictions because more stacking predictions, which will all undergo the online update procedure, provide the online update with more options to choose from. Fig 3 illustrates the training process of stacking.

### 4.3 Bootstrap aggregation

In order to reduce the model variance, we further implement bootstrap aggregation [44]. Assume our original training set Z has B bootstrap samples $Z^{*b}$, b=1,2,...,B, and we use each bootstrap sample to construct a classifier with its prediction result being $\hat{f}^{*b}(x)$. The final prediction after the bootstrap aggregation is the average: $\hat{f}_{bag}(x) = \frac{1}{B}\sum_{b=1}^{B}\hat{f}^{*b}(x)$.

We set the sample size to be 80% of the original training set and create 10 bootstrap samples. We use each sample to train the base and ensemble models. The final stacking predictions, which will then pass on to online update, are the each sample's bootstrap aggregated stacking predictions. Fig 4 shows the bootstrap aggregation process.

### 4.4 Online update strategy

Before diving into the details of online update, we first explain why it provides the dynamic flexibility that constitutes a key feature of dynABE. Machine learning models generally use the training data to uncover an underlying pattern between the input and output variables. This method is powerful for tasks like image classification when the pattern to discover is

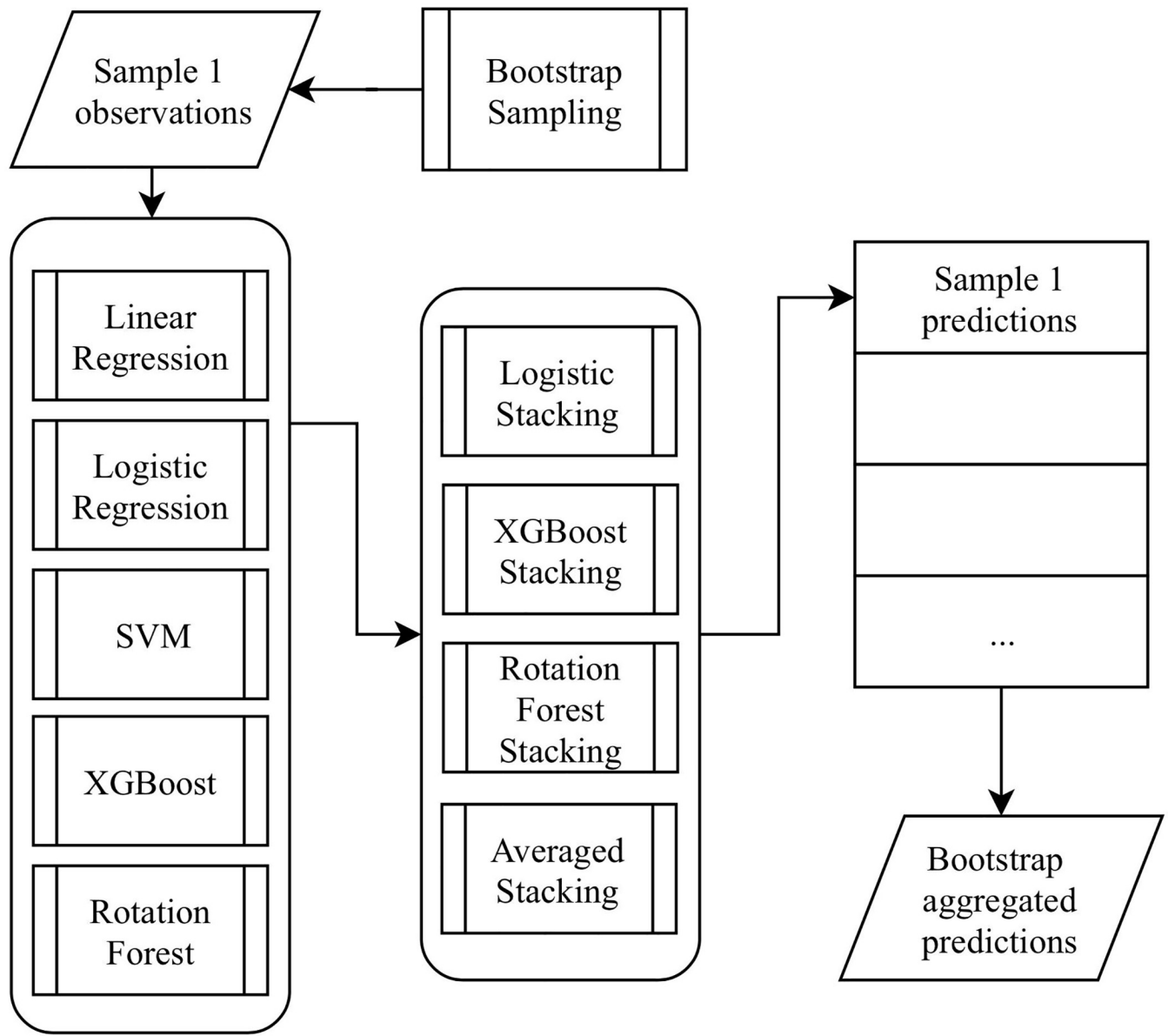

**Fig 4. Stabilizing prediction results through bootstrap aggregation.** We show processing sample 1 as an example. The same process would be repeated 10 times in practice.

https://doi.org/10.1371/journal.pone.0212487.g004

unchanged: a cat always has one tail, four feet, and two fluffy ears. However, the stock market is dynamic, and its price patterns can change with time. Directly constructing machine learning models based on the training time period can ignore some important factors. For example, assume we know that the commodity market has to influence the stock price of the producer company in general, but it does not have to during the training period. In a regular machine learning model, this information is thus filtered out. However, once the price pattern changes and the commodity market starts to influence the stock price again in the active trading period, the model will be inaccurate because it has already filtered out the commodity market data.

This problem calls for the need of updating the model according to changing price patterns during the active trading period, which is less intuitive from simply retraining the model every day with new data. Adding one more day of data to the hundreds of days of training set barely affects the model, is computationally expensive, and makes the model prone to overfitting. For most time-series models, including dynABE, retraining the entire model with new data is necessary after a *long* period of time, such as a year, and we will give a sense of how to determine when dynABE needs to be retrained in the Results and analysis section. However, here we are only interested in updating in the short-term to increase accuracy.

Online update in dynABE handles this problem. After we identify general factors that influence the stock price and encode them as advisors, these factors are readily available for the model at all times and will not be filtered out. dynABE can then weigh these factors during the active trading period and change the weights according to price patterns it observes. It is important to point out that online update deliberately only affects the model at the top level of advisors and does not change the base models. This avoids the dilemma of retraining every day and adds robustness to the model.

Now we discuss the formulation of online update. Using the ensemble learning framework described in the previous section, each advisor would make four predictions on the validation set, as shown in Fig 5. Since there are three advisors, we have a total of 12 predictions, which we will term as "agents" of the online update process.

In the online update strategy, the first important parameter is the update frequency $f$. Intuitively, $f$ is the length of the update window where each agent's weight is updated every $f$ days mostly determined by the performances in the past $f$ days. $f$ should be small enough to swiftly update the weights, yet it also should be large enough to capture meaningful relationships between the agents before moving on to the next update.

The second important parameter, decay rate $\gamma$, determines how much we want to discount the influence of previous update windows. Each time when we calculate the score of the current update window, we add it to the scores of the previous update windows, where the previous scores are all multiplied by $\gamma$ each time. $\gamma$ is in the range of 0 and 1, where a $\gamma$ of 0 is when we do not consider days prior to the update window at all, and a $\gamma$ of 1 is when we consider days prior to the update window equally as important as days in the update window. $\gamma$ exponentialy decays and converges the influence of old update windows, and it is analogous to the temporal-difference method [48], which deals with continually taking the means of temporally successive data. The specific definition of decay rate in formula is presented later.

The third important parameter is diversity bias $\lambda$. Base model diversity is important for successful model ensembles. For dynABE, because each advisor has a different feature set, they should already make diverse predictions. Nevertheless, we use diversity bias when we want to further encourage model diversity. Intuitively, $\lambda$ represents how much we want to award the case when one agent makes a correct prediction while other agents are wrong. We implement it as a quality measure of correct predictions: when an agent is correct on one day, the more the other agents make mistakes on that day, the higher the quality this correct prediction has.

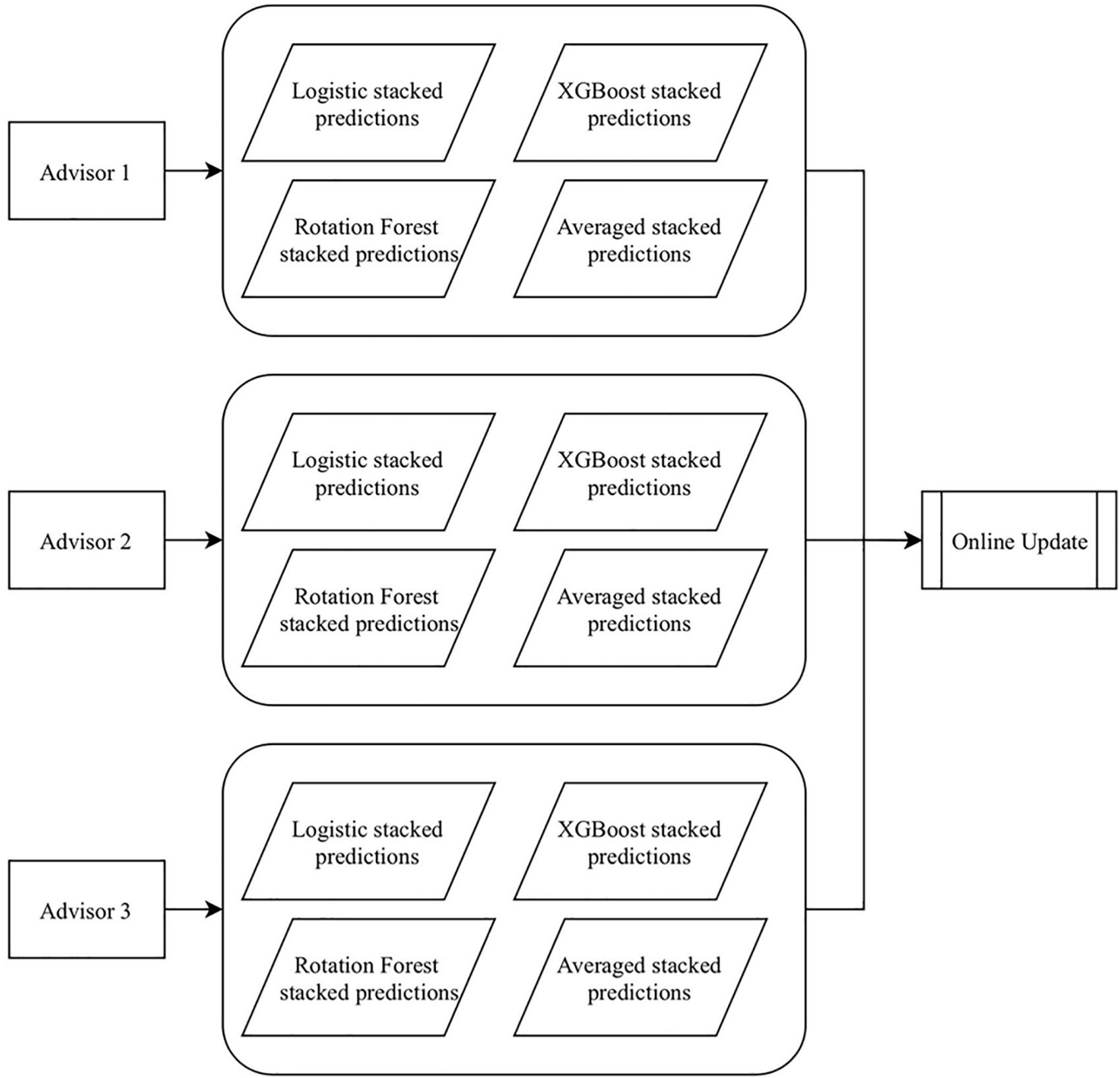

**Fig 5. Available agents to be passed to online update.**

https://doi.org/10.1371/journal.pone.0212487.g005

With a high $\lambda$, the agent that makes the highest quality of correct predictions often wins over the agent that makes the highest quantity of correct predictions.

To summarize, we update the weights for each agent based on previous performances, determined by update frequency $f$ and decay rate $\gamma$. Diversity bias $\lambda$ further adjusts the weights. Now we present our implementation. Assume that we are currently in an online update window $\tau_i$ of $f$ days, and there are $N$ agents in total, $\boldsymbol{a} = (a_1, a_2, \ldots, a_N)$. An agent $a_n$'s score on the $j^{th}$ day, $S_{a_n}^{j}$, is updated in the following fashion until this update window is over:

$$S_{a_n}^{j} = S_{a_n}^{j-1} + \frac{\phi(a_n, j)}{1 + \lambda \sum \phi(a_{-n}, j)} \quad (1)$$

where

$$\phi(a_n, j) = \begin{cases} 1, & \text{if } a_n \text{ makes a correct prediction on the } j^{th} \text{ day} \\ 0, & \text{otherwise} \end{cases}$$

$a_{-n}$ denotes all advisors excluding $a_n$, and $\lambda$ is the diversity bias. Therefore, if agent $a_n$ makes a correct prediction on the $j^{th}$ day, the more the other agents are correct on that day, the higher $\Sigma \phi(a_{-n}, i)$ is, and the smaller the score of $a_n$ will be increased. After the update window is over, denote the accumulated score for $a_n$ in this window to be $S_{a_n}^{window}$. Then $a_n$'s final score after this update window, $S_{a_n}^{\tau_i}$, is calculated as:

$$S_{a_n}^{\tau_i} = S_{a_n}^{window} + \gamma S_{a_n}^{\tau_{i-1}} \quad (2)$$

where $\gamma$ is the decay rate between 0 and 1. $a_n$'s new weight is finally calculated by normalizing the score to a 0 to 1 scale:

$$w_{a_n} = \frac{S_{a_n}^{\tau_i}}{\sum_{j=1}^{N} S_{a_j}^{\tau_i}} \quad (3)$$

The entire online update process is illustrated by the following algorithms:

Algorithm 1: *Online update with diversity bias*

**Inputs:**
Predictions by N agents as matrix $\hat{Y}_a = (\hat{y}_{a_1}, \hat{y}_{a_2}, \ldots, \hat{y}_{a_N})$ where each column is $\hat{y}_{a_n}$, a vector of predictions made by agent $a_n$
Actual outcome: *y*
Update frequency: *f*
Decay rate: *γ*
Diversity bias: *λ*
**Output:**
Predictions after online update: $\hat{y}_{final}$

1. Initialize:
   current day i = 0
   $\hat{y}_{final} = <>$, an empty vector
   score vector $\boldsymbol{S}^1 = <S_{a_1}^{1}, S_{a_2}^{1}, \ldots, S_{a_N}^{1}> = \vec{0}$, a zero vector
   weight vector $\boldsymbol{w}^1 = <w_{a_1}^{1}, w_{a_2}^{1}, \ldots, w_{a_N}^{1}> = \frac{\vec{1}}{N}$, a vector whose values are all $\frac{1}{N}$
2. While i + f < length(*y*), do:
   a. Define update window: $\boldsymbol{\tau} = i(i + f)$
   b. Take the rows of matrix $\hat{Y}_a$ indexed by $\boldsymbol{\tau}$ as $\hat{Y}_a^{\tau} = \hat{Y}_a[\boldsymbol{\tau},:]$, and take the elements of *y* indexed by $\boldsymbol{\tau}$ as $\boldsymbol{y}^{\tau} = \boldsymbol{y}[\boldsymbol{\tau}]$
   c. Make predictions for the next f days: $\hat{y}_{new} = \hat{Y}_a^{\tau} \boldsymbol{w}^i$
   d. Append $\hat{y}_{new}$ to $\hat{y}_{final}$
   e. Update weights: $\boldsymbol{w}^{i+f} = update\_weights(\hat{Y}_a^{\tau}, \boldsymbol{y}^{\tau}, \boldsymbol{S}^i, \gamma, \lambda)[0]$
   f. Record scores: $\boldsymbol{S}^{i+f} = update\_weights(\hat{Y}_a[T,:], \boldsymbol{y}[T], \boldsymbol{S}^i, \gamma, \lambda)[1]$
   g. Update day: i(i + f)
3. If i < length(*y*), then:
   a. Process the leftover days: $\hat{y}_{new} = \hat{Y}_a[i:length(y),:]\boldsymbol{w}^i$
   b. Append $\hat{y}_{new}$ to $\hat{y}_{final}$

```
        4. Output: ŷ_final
Algorithm 2: update_weights
Inputs:
Predictions by N agents for update window τ as matrix Ŷ_a^τ =
(ŷ_{a_1}^τ, ŷ_{a_2}^τ, ..., ŷ_{a_N}^τ) where each column is ŷ_{a_n}^τ, a vector of predictions made
by agent a_n in update window τ
Actual outcome for update window τ: y^τ
Old score vector: S^old
Decay rate: γ
Diversity bias: λ
Output:
Updated weight vector: w' = (w_{a_1}, w_{a_2}, ..., w_{a_N})
Updated score vector: S' =< S_{a_1}, S_{a_2}, ..., S_{a_N} >
    1. Initialize score: S' = 0⃗
    2. For day i in 1:length(y^τ):
        a. For agent a_n in {a_1, a_2, ...a_N}, update:
```

$$S_{a_n} = S_{a_n} + \frac{\phi(a_n, i)}{1 + \lambda \sum \phi(a_{-n}, i)} \quad \phi(a_n, i) = \begin{cases} 1, & \text{if } \hat{y}'_{a_k}[i] = y'[i] \\ 0, & \text{otherwise} \end{cases}$$

```
        3. Calculate final score vector: S' = S' + γS^old
        4. Normalize scores to obtain weights:
            a. sum_scores = S_{a_1} + S_{a_2} + ... + S_{a_N}
            b. If sum_scores == 0, then w' = (1/N, 1/N, ..., 1/N)  // None of the agents
made correct predictions
            c. Else, w' = S'/sum_scores
        5. Output: (w', S')
```

For the sake of simplicity, we fix the decay rate to be 0.8 and only choose values for update frequency and diversity bias, but all three parameters can be tuned together, which is a potential to further improve model performance.

## 5 Results and analysis

We test dynABE on three cobalt-related companies: Jinchuan Group, Sumimoto Metal Mining, and Zijin Mining. As previously mentioned, we create three advisors for this case study. We formulate Advisor 1 to represent macroeconomic factors, Advisor 2 to represent the cost of production, and Advisor 3 to represent the metal market. We will refer to them as Advisor 1, 2, and 3 for the rest of this section.

We evaluate dynABE on the classification accuracy of each advisor as well as the accuracy after all the advisors are combined in online update. Then we plot the weight update histories to visualize online update at work. Afterward, we discuss an interesting observation of accuracy decay, which is related to choosing the length of active trading before dynABE is updated with new data. Finally, we use predictions of dynABE for creating a naïve trading strategy to show the model's financial value.

### 5.1 Individual advisor performances

We use misclassification errors as the evaluation metric, which is the percentage of the trading days for which the models predicts the wrong stock trend. We first look at each advisor's base and ensemble models' errors on validation sets before online update. Since we use bootstrap sampling to stabilize the base models, we present the average errors over all 10 samples for the

Table 2. Misclassification errors of each advisor for all three companies during the validation period. The best performance of each advisor is bolded.

| Company | Classifier | Advisor 1 Error | Advisor 2 Error | Advisor 3 Error |
|---|---|---|---|---|
| Jinchuan | Linear Regression | (average) 36.33% | (average) 43.04% | (average) 39.74% |
| | Logistic Regression | (average) 39.00% | (average) 40.29% | (average) 39.45% |
| | SVM | (average) 35.72% | (average) 40.47% | (average) 38.79% |
| | XGBoost | (average) 34.86% | (average) 36.35% | **(average) 35.28%** |
| | Rotation forest | (average) 40.00% | (average) 41.84% | (average) 39.13% |
| | Logistic Stacking | 35.43% | **33.86%** | 36.75% |
| | XGBoost Stacking | **34.38%** | 33.86% | 35.70% |
| | Rotation Forest Stacking | 35.43% | 38.06% | 38.85% |
| | Averaged Stacking | 35.70% | 34.12% | 36.75% |
| Sumitomo | Linear Regression | (average) 34.03% | (average) 44.22% | (average) 40.06% |
| | Logistic Regression | (average) 36.06% | (average) 43.72% | (average) 38.28% |
| | SVM | (average) 35.75% | (average) 43.75% | (average) 40.17% |
| | XGBoost | (average) 36.81% | **(average) 43.31%** | (average) 35.22% |
| | Rotation forest | (average) 35.89% | (average) 43.89% | (average) 39.78% |
| | Logistic Stacking | 31.67% | 44.17% | 35.00% |
| | XGBoost Stacking | 32.22% | 45.28% | 36.11% |
| | Rotation Forest Stacking | 32.50% | 43.33% | 34.72% |
| | Averaged Stacking | **31.94%** | 43.89% | **34.17%** |
| Zijin | Linear Regression | (average) 43.14% | (average) 43.72% | (average) 42.31% |
| | Logistic Regression | (average) 41.86% | (average) 43.25% | (average) 42.36% |
| | SVM | (average) 43.06% | (average) 43.28% | (average) 41.31% |
| | XGBoost | (average) 41.81% | (average) 44.61% | (average) 42.19% |
| | Rotation forest | **(average) 41.33%** | (average) 44.58% | (average) 43.00% |
| | Logistic Stacking | 42.78% | **41.67%** | 42.50% |
| | XGBoost Stacking | 42.50% | 42.50% | 40.83% |
| | Rotation Forest Stacking | 42.50% | 43.89% | **40.00%** |
| | Averaged Stacking | 41.94% | 42.78% | 41.11% |

https://doi.org/10.1371/journal.pone.0212487.t002

base models. Table 2 shows the misclassification errors on the validation set of all three advisors before online update.

The best performance of each advisor for each company is bolded. For Jinchuan, Advisor 2 achieves the best performance with a 33.86% misclassification error with logistic stacking. For Sumitomo, Advisor 1 performs best at a 31.94% error with averaged stacking while Zijin shows that Advisor 3 is most accurate with a 40.00% error with rotation forest stacking. No single base model consistently outperforms the others in all three datasets, which justifies our decision to use an ensemble-based approach. In general, stacking succeeds at improving the base model performances, such as reducing the error from 34.03% to 31.94% in Advisor 1 of Sumitomo. Nevertheless, stacking does not always outperform every single base model. This situation happens when only one base model is very strong while others are weak, such as Advisor 3 of Jinchuan for which the base model XGBoost outperforms all the stacking models.

## 5.2 Online update performances

Online update predictions are evaluated differently to take the initialization period into consideration. For an update frequency of $f$, we evaluate online update by excluding its predictions of the first $f$ days, for which we use to assign the initial weights to the advisors.

**Table 3. Online update experiments with Jinchuan.** Here we show common hyperparameter combinations and their effects on online update's misclassification error. Then we grid search on the validation set and present the searched optimal combinations.

**Company: Jinchuan**

| Update Frequency | Diversity Bias | Error |
|---|---|---|
| 3 | 0 | 33.86% |
| 3 | 1 | 33.86% |
| 3 | 10 | 33.60% |
| 5 | 0 | 33.51% |
| 5 | 1 | 33.24% |
| 5 | 10 | 32.71% |
| 10 | 0 | 32.88% |
| 10 | 1 | 32.35% |
| 10 | 10 | 32.35% |
| (grid search) 5 | (grid search) 5 | 31.12% |

https://doi.org/10.1371/journal.pone.0212487.t003

**Table 4. Online update experiments with Sumitomo.**

**Company: Sumitomo**

| Update Frequency | Diversity Bias | Error |
|---|---|---|
| 3 | 0 | 31.93% |
| 3 | 1 | 31.93% |
| 3 | 10 | 33.61% |
| 5 | 0 | 32.39% |
| 5 | 1 | 31.83% |
| 5 | 10 | 34.08% |
| 10 | 0 | 31.71% |
| 10 | 1 | 32.29% |
| 10 | 10 | 34.00% |
| (grid search) 12 | (grid search) 0 | 31.61% |

https://doi.org/10.1371/journal.pone.0212487.t004

**Table 5. Online update experiments with Zijin.**

**Company: Zijin**

| Update Frequency | Diversity Bias | Error |
|---|---|---|
| 3 | 0 | 42.58% |
| 3 | 1 | 42.58% |
| 3 | 10 | 40.34% |
| 5 | 0 | 42.54% |
| 5 | 1 | 41.41% |
| 5 | 10 | 40.28% |
| 10 | 0 | 42.29% |
| 10 | 1 | 41.14% |
| 10 | 10 | 40.86% |
| (grid search) 40 | (grid search) 31 | 37.19% |

https://doi.org/10.1371/journal.pone.0212487.t005

Since decay rate is always fixed at 0.8, we only need to set the values of update frequency and diversity bias. Here we experiment with several common values for these two hyperparameters as a guideline. 3 to 10 is a common range for the update frequency and 0 to 10 for

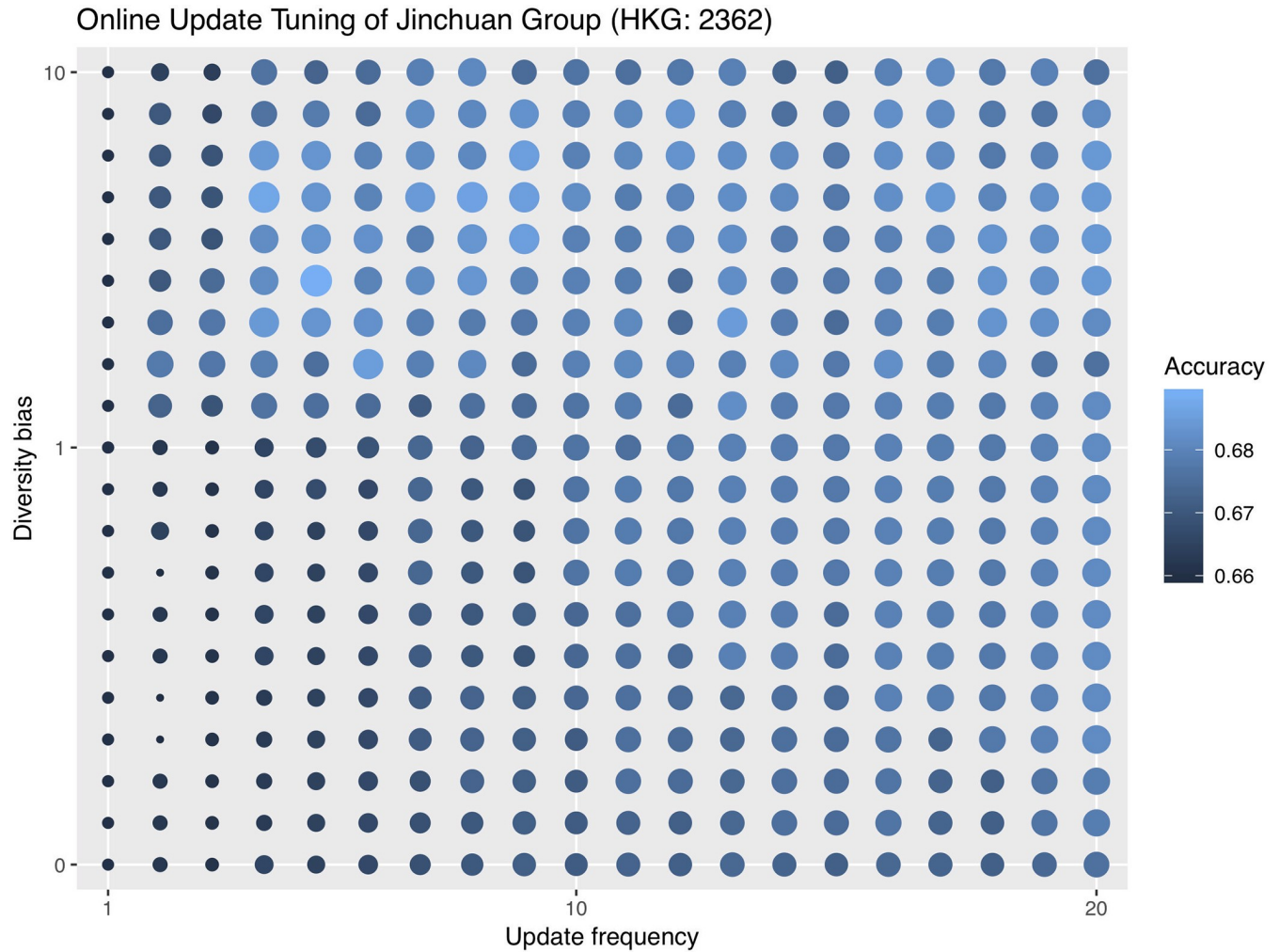

**Fig 6. Online update hyperparameter grid search for Jinchuan.** Bigger and brigher bubbles represent higher accuracies.

https://doi.org/10.1371/journal.pone.0212487.g006

diversity bias. Therefore, we try update frequencies of 3, 5, and 10 and diversity biases of 0, 1, 10 and record the respective errors for each combination. We do not develop a hyperparameter tuning strategy for online update in this paper, but in order to aid future developments of tuning strategies, we do an exhaustive hyperparameter grid search on the validation set in addition to the common values. Tables 3–5 show each company's online update misclassification errors first with common hyperparameter combinations and, in the last row, with the grid searched optimal hyperparameter combinations.

With the exception of Zijin, whose optimal hyperparameter combination is way outside the normal range, the performances with common hyperparameter combinations do not differ significantly from the optimal performances.

We visualize the hyperparameter grid search with a bubble plot. It is important to point out that this is not a tuning process because we are using the validation set. Instead, the grid search experiment is for an intuition of what hyperparameter combinations are often more effective. Without a tuning strategy, one is not always able to choose the optimal hyperparameter combination. Figs 6–8 are the hyperparameter grid search plots for all three companies. Bigger and brighter bubbles represent higher accuracies, corresponding to the legend. The

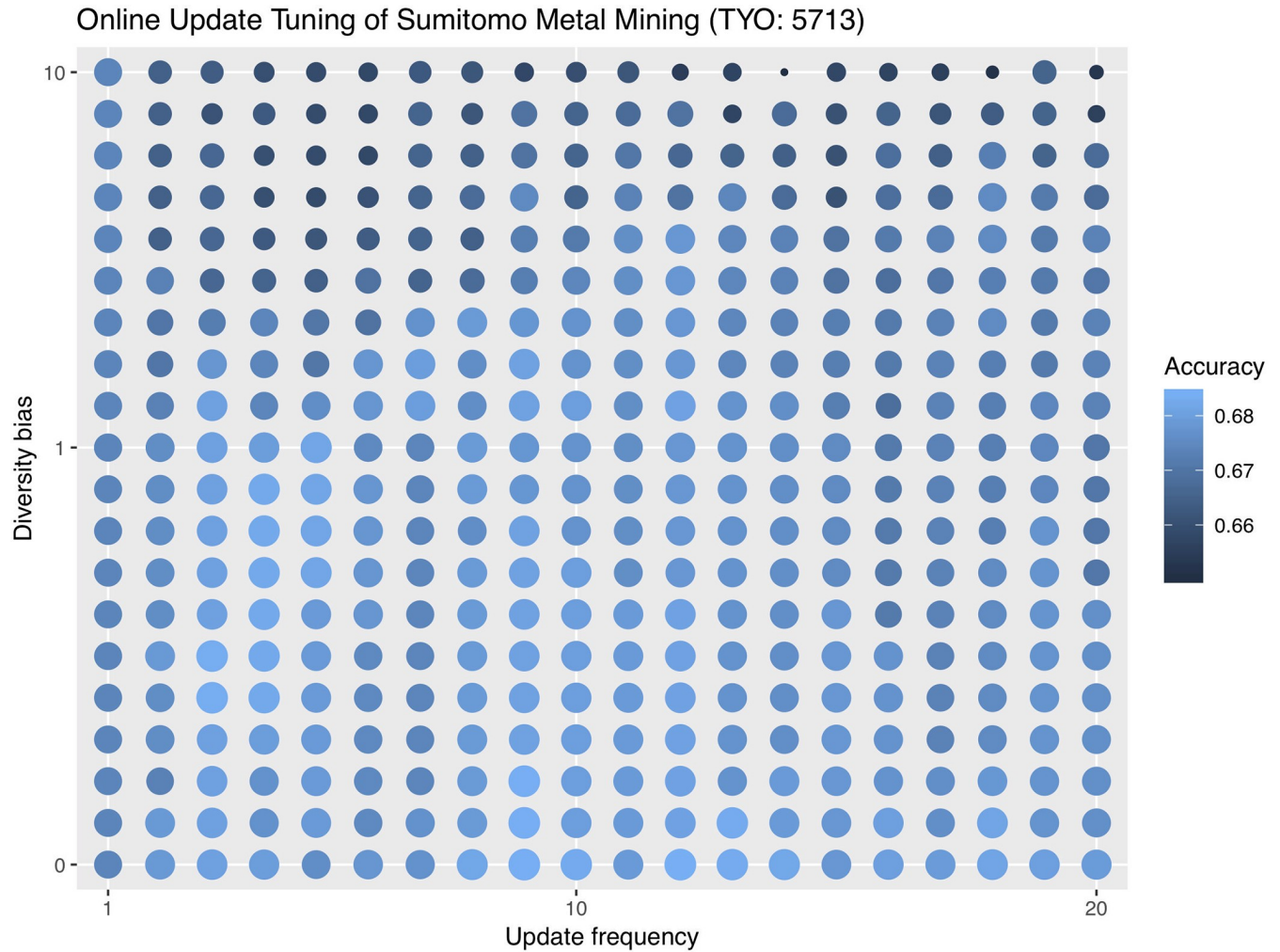

**Fig 7. Online update hyperparameter grid search for Sumitomo.**

https://doi.org/10.1371/journal.pone.0212487.g007

tuning range for Zijin Mining is bigger than that for Jinchuan and Sumitomo so the bubbles are denser.

For Jinchuan, the patterns are the most obvious as high accuracy bubbles are usually in the upper right portion of the plot. This means that a high diversity bias combined with a big update frequency is likely to yield good results. On the other hand, low diversity bias with small update frequency should be avoided. Sumitomo is not as sensitive to different hyperparameter combinations, and as long as the diversity bias is not too high, such as between 0 and 1, accuracies are high regardless of update frequency. We later show that Sumitomo also does not benefit much from online update, so this might be why different hyperparameter combinations do not affect Sumitomo's performance dramatically. Last but not least, Zijin's optimal hyperparameter combinations are found in a more extreme range than the other two companies, so we show a more exhaustive tuning plot. Zijin generally favors very high diversity biases, mostly above 20, coupled with relatively big update frequencies.

We now compare the stacking errors with the online update errors achieved with the optimal hyperparameter combinations, which we consider to be expected optimal. Table 6 shows

PLOS ONEDynamic Advisor-Based Ensemble (dynABE) for stock prediction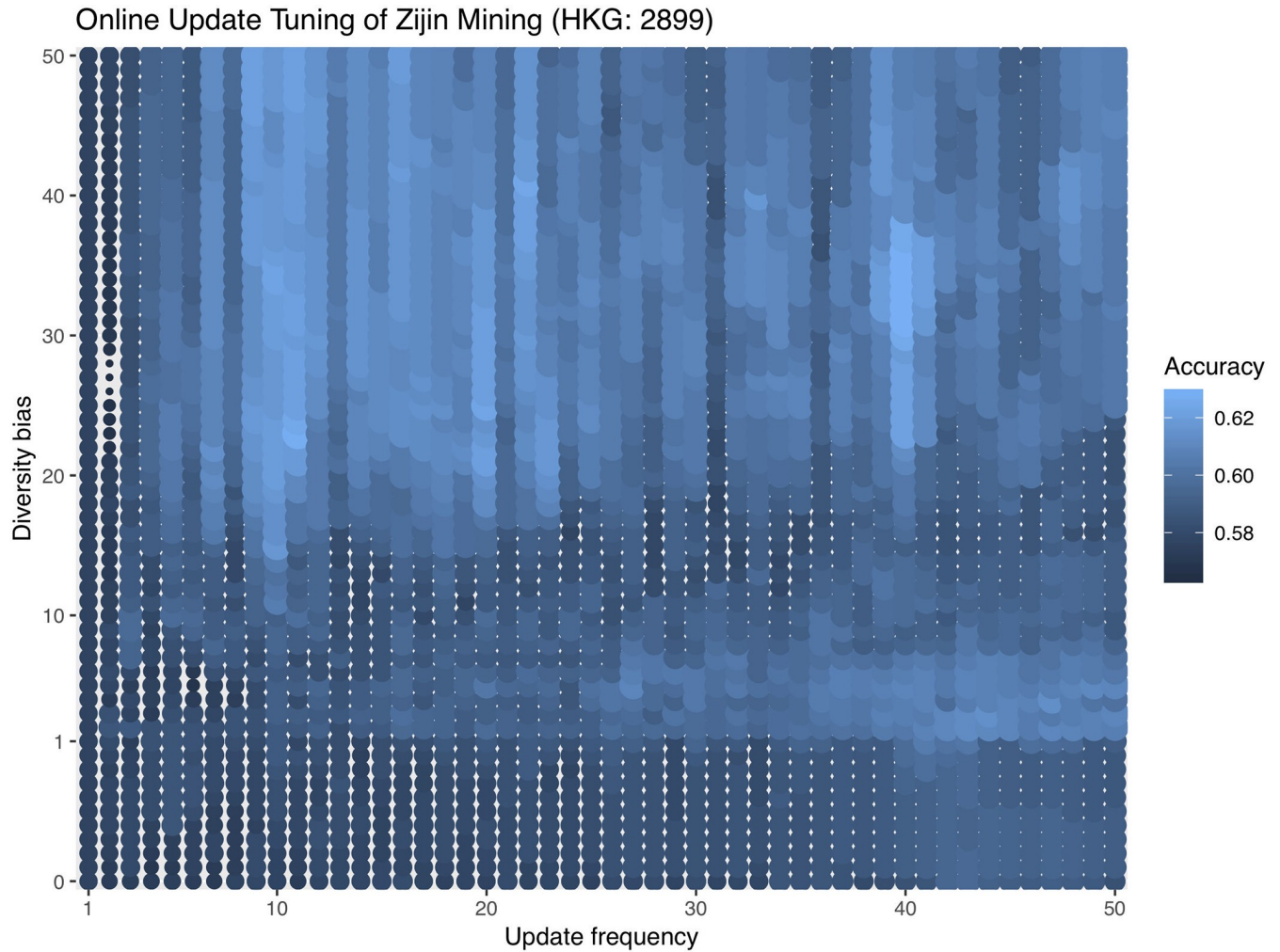

**Fig 8. Online update hyperparameter grid search for Zijin.** Bubbles are denser for Zijin because its optimal hyperparameter combinations are outside the normal range, so its tuning range is also greater.

https://doi.org/10.1371/journal.pone.0212487.g008

**Table 6. Comparison of stacking and online update errors.** The best performance of each company is bolded.

| Company | Advisor 1 Stacking Error | Advisor 2 Stacking Error | Advisor 3 Stacking Error | Online Update Error |
|---|---|---|---|---|
| Jinchuan | 35.43% (Logistic Stk.) | 33.86% (Logistic Stk.) | 36.75% (Logistic Stk.) | **31.12%** (expected optimal) |
| | 34.38% (XGBoost Stk.) | 33.86% (XGBoost Stk.) | 35.70% (XGBoost Stk.) | |
| | 35.43% (Rot. Forest Stk.) | 38.06% (Rot. Forest Stk.) | 38.85% (Rot. Forest Stk.) | |
| | 35.70% (Averaged Stk.) | 34.12% (Averaged Stk.) | 36.75% (Averaged Stk.) | |
| Sumitomo | 31.67% (Logistic Stk.) | 44.17% (Logistic Stk.) | 35.00% (Logistic Stk.) | **31.61%** (expected optimal) |
| | 32.22% (XGBoost Stk.) | 45.28% (XGBoost Stk.) | 36.11% (XGBoost Stk.) | |
| | 32.50% (Rot. Forest Stk.) | 43.33% (Rot. Forest Stk.) | 34.72% (Rot. Forest Stk.) | |
| | 31.94% (Averaged Stk.) | 43.89% (Averaged Stk.) | 34.17% (Averaged Stk.) | |
| Zijin | 42.78% (Logistic Stk.) | 41.67% (Logistic Stk.) | 42.50% (Logistic Stk.) | **37.19%** (expected optimal) |
| | 42.50% (XGBoost Stk.) | 42.50% (XGBoost Stk.) | 40.83% (XGBoost Stk.) | |
| | 42.50% (Rot. Forest Stk.) | 43.89% (Rot. Forest Stk.) | 40.00% (Rot. Forest Stk.) | |
| | 41.94% (Averaged Stk.) | 42.78% (Averaged Stk.) | 41.11% (Averaged Stk.) | |

https://doi.org/10.1371/journal.pone.0212487.t006

Table 7. **Comparison between baseline models and dynABE on the validation set.** Here we use misclassification errors as the evaluation metric. The best baseline performances are italicized, and the best overall performances are bolded.

| Company | Support Vector Machine | 3-layer Neural Network | Random Forest | dynABE |
|---|---|---|---|---|
| Jinchuan | 37.53% | 35.43% | *34.12%* | **31.12%** |
| Sumitomo | 41.39% | 44.44% | *38.06%* | **31.61%** |
| Zijin | 43.61% | 40.28% | *39.72%* | **37.19%** |

https://doi.org/10.1371/journal.pone.0212487.t007

the expected optimal performance boost from stacking to online update. The errors correspond to predicions made on the validation sets.

Online update outperforms all stacking predictions from three advisors for all the companies. Specifically, Jinchuan and Zijin both benefit significantly from online update. However, online update does not improve Sumitomo's best stacking predictions as much, where its best stacking error is 31.67% and the best online update error is 31.61%. This might explain why Sumitomo is less sensitive to different hyperparameter combinations during grid search in the previous experiment. We believe that Sumitomo benefits less from online update because its

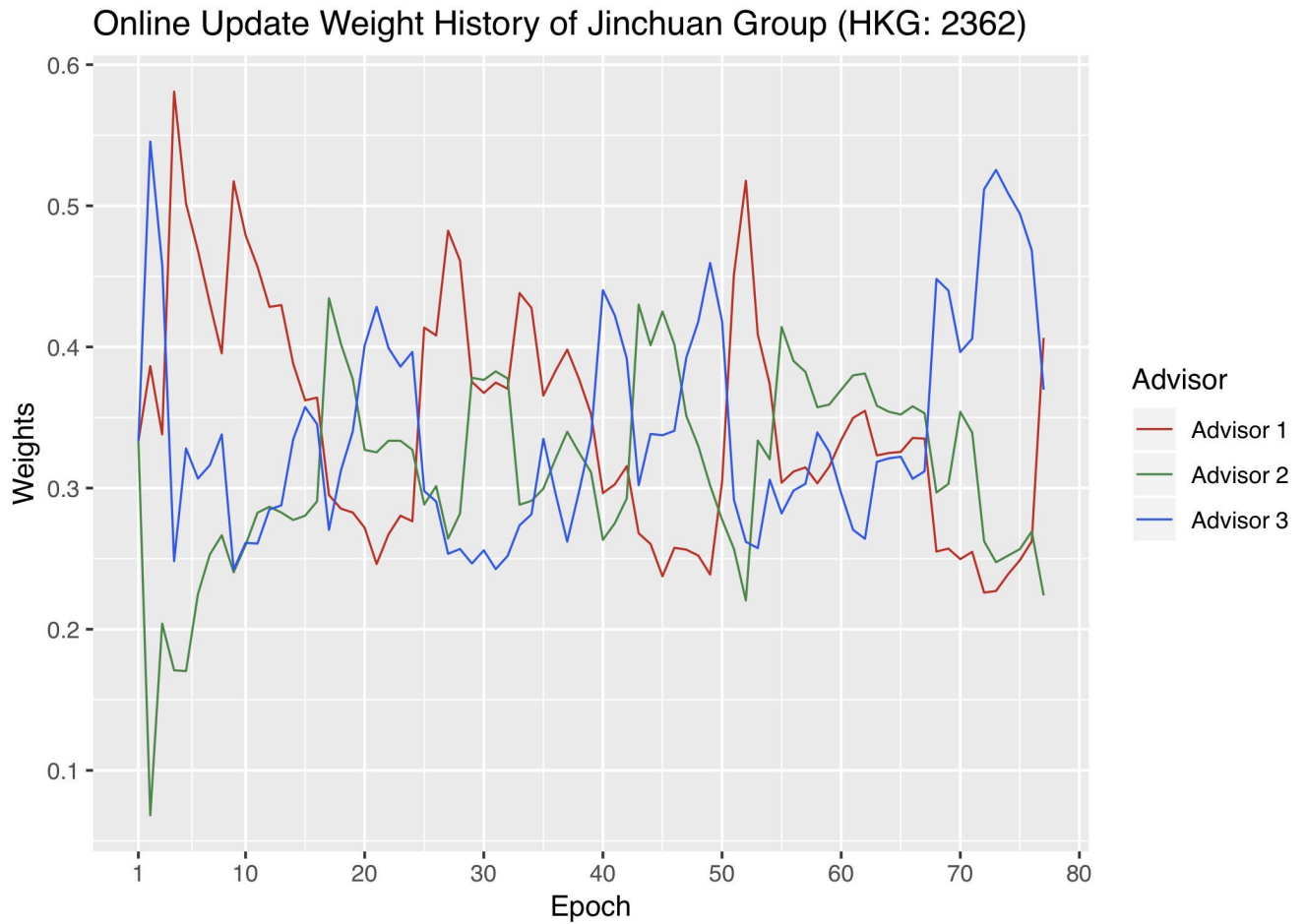

**Fig 9. Advisor weight history of Jinchuan.** Different advisors are represented by different colors, corresponding to the legend. The x-axis is the epochs. Each epoch means one weight update. The y-axis is the weights. A higher weight means that a certain advisor plays a more important role during online update.

https://doi.org/10.1371/journal.pone.0212487.g009

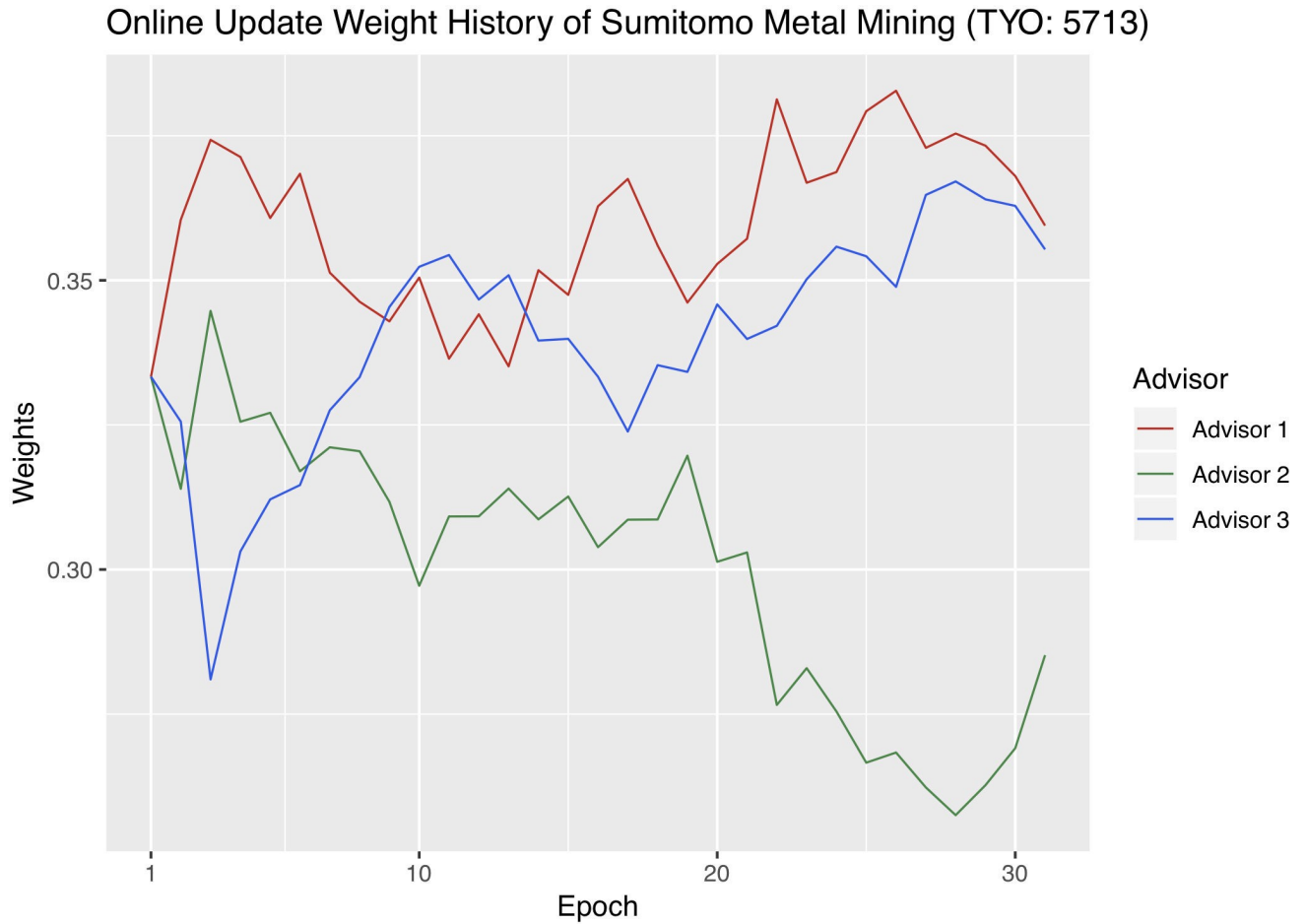

**Fig 10. Advisor weight history of Sumitomo.**

https://doi.org/10.1371/journal.pone.0212487.g010

Advisor 2 is significantly weaker than the other two advisors, so it cannot contribute much to compensating the other two advisors' mistakes during online update.

### 5.3 Comparison with baseline models

Here we compare the classification accuracies of dynABE with three baseline models for stock prediction: support vector machine, neural network, and random forest. These are popular models in the field for stock prediction as discussed in the Related works section. In order to make a fair comparison of the models' predictive abilities only, we use the same feature selection process for the baseline models as dynABE. All baseline models' hyperparameters are tuned extensively with grid search. For neural network, we use a 3-layer structure with 20 hidden nodes, optimized with stochastic gradient descent with momentum. Table 7 compares the classification accuracies between baseline models and dynABE.

Among the three baseline models, random forest consistently outperforms support vector machine and neural network. This result is consistent with comparative studies in the field, such as Patel *et al.*'s work [17] and Ballings *et al.*'s work [18] discussed before, which show that random forest is one of the best performing models for stock prediction. Nevertheless, dynABE consistently outperforms all the baseline models in all three case studies by a noticeable degree. In addition, besides Zijin, which tends to favor more extreme online update

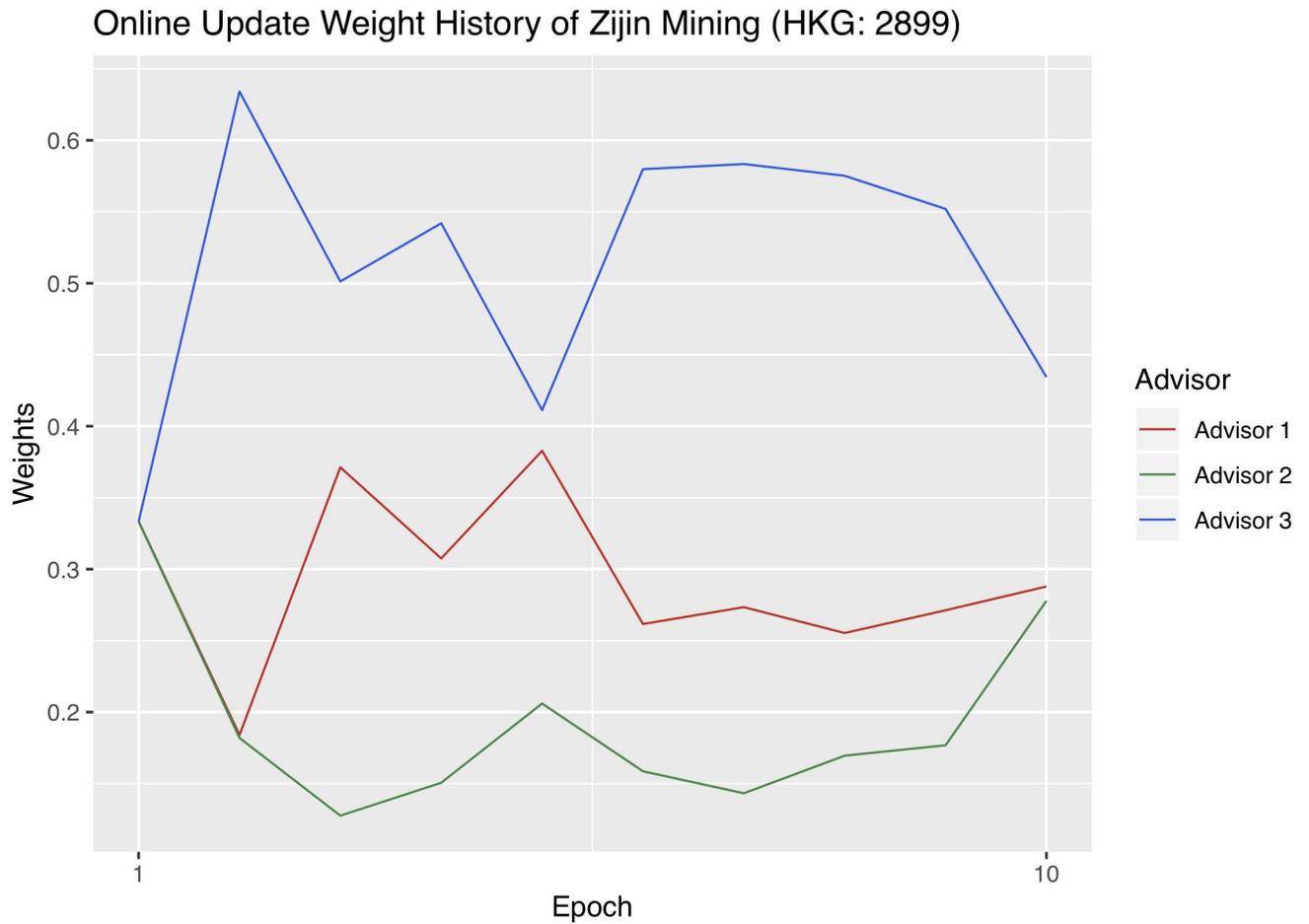

**Fig 11. Advisor weight history of Zijin.**

https://doi.org/10.1371/journal.pone.0212487.g011

hyperparameter combinations as shown in the previous section, dynABE with any common hyperparameter combinations is able to outperform all the tuned baseline models (see Tables 3–5).

### 5.4 Weight update histories in online update

In order to better understand online update, we visualize the each advisor's weight changes with each company's optimal hyperparameter combinations. We sum the weights of the four stacking methods that belong to the same advisor into one advisor weight and plot the weights of the three advisors. Figs 9–11 are the weight history plots. The x-axis is the epoch, and each epoch means one weight update. The y-axis represents the weights.

Recall that Advisor 1 represents macroeconomics, Advisor 2 the cost of production, and Advisor 3 the metal market. The advisor that receives the highest weight at a certain epoch affects the result of online update the most. The weights of the three advisors always add up to one.

Jinchuan has the most balanced weight update history because none of the three advisors consistently receives bigger weights than the other two. Zijin has the most severe advisor imbalance as Advisor 3 always receives the biggest weights, and 2 the smallest, throughout the validation period. In addition, since Sumitomo's Advisor 2 was shown to be noticeably weaker

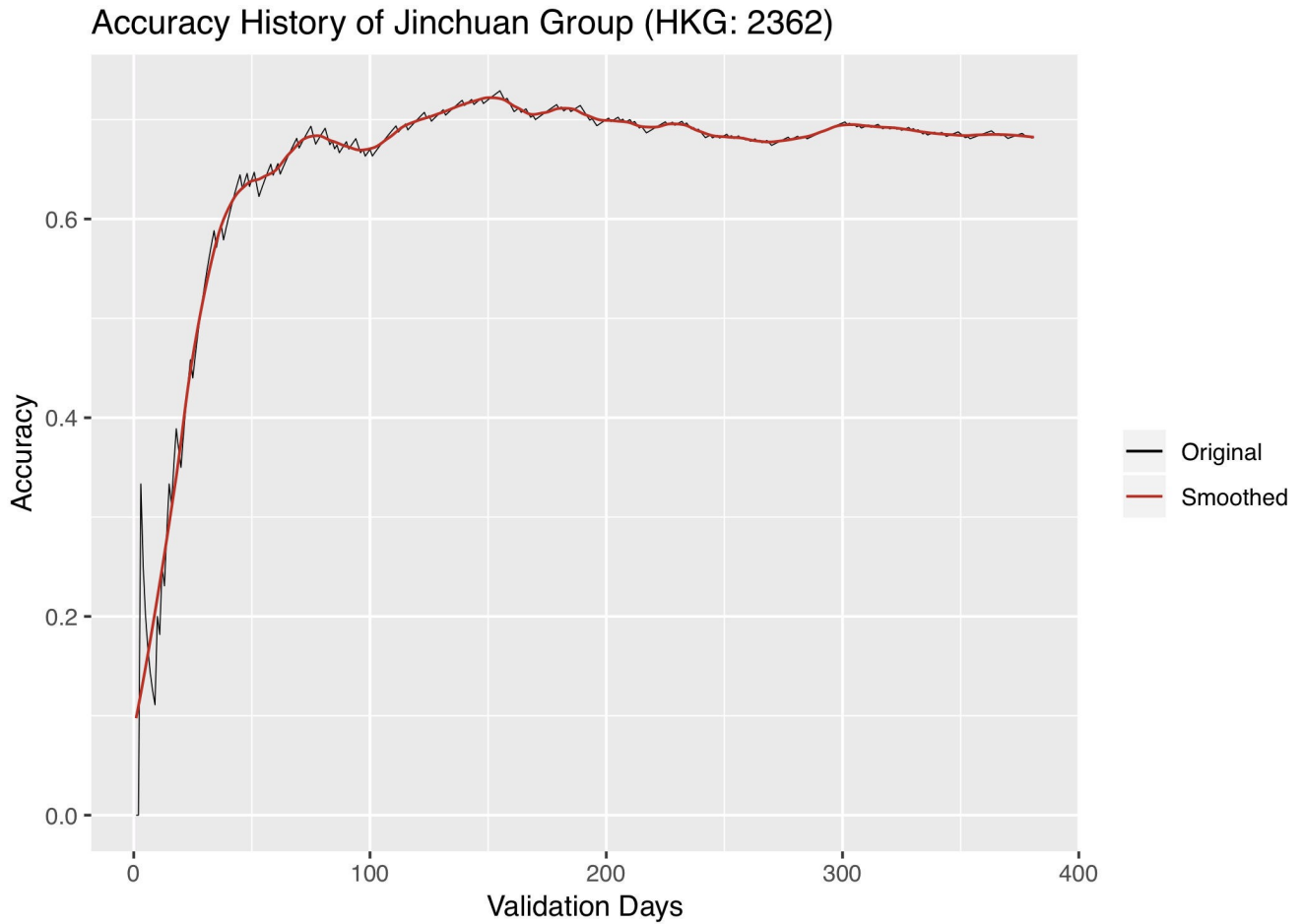

**Fig 12. Accuracy history of Jinchuan.** x-axis is the dates, and y-axis is the cumulative accuracy up to a certain day.

https://doi.org/10.1371/journal.pone.0212487.g012

than Advisor 1 and 3 before in the Individual advisor performances section, it is reasonable that Advisor 2, plotted in green in Fig 10, continuously receives smaller weights throughout the trading period.

### 5.5 Accuracy gain and decay in online update

We further experiment with changing the length of validation, i.e. the active trading period, and observing its effect on classification accuracy. We carry out the following experiment: a dynABE model trained on the training set starts with the first day of the validation set, performs online update, and records the classification accuracy for the first day; then we add one more day and record the cumulated accuracy for the first two days; we repeat this process until we include all days of the validation period. Figs 12–14 are the accuracy history plots of this experiment. An additional smoothed trend line is plotted to better visualize the trend.

Interestingly, we observe a universal pattern of a steady accuracy gain at first. It is sometimes followed by a steady accuracy decay after the highest accuracy is reached, most obviously in Sumitomo. We believe that the initial accuracy gain as the length of validation gets further into the future is a sign of the online update learning more accurate relationships between the advisors, after seeing more historical data. On the other hand, the accuracy decay might be due to the expiration of patterns each advisor observes from its training set, so the online update

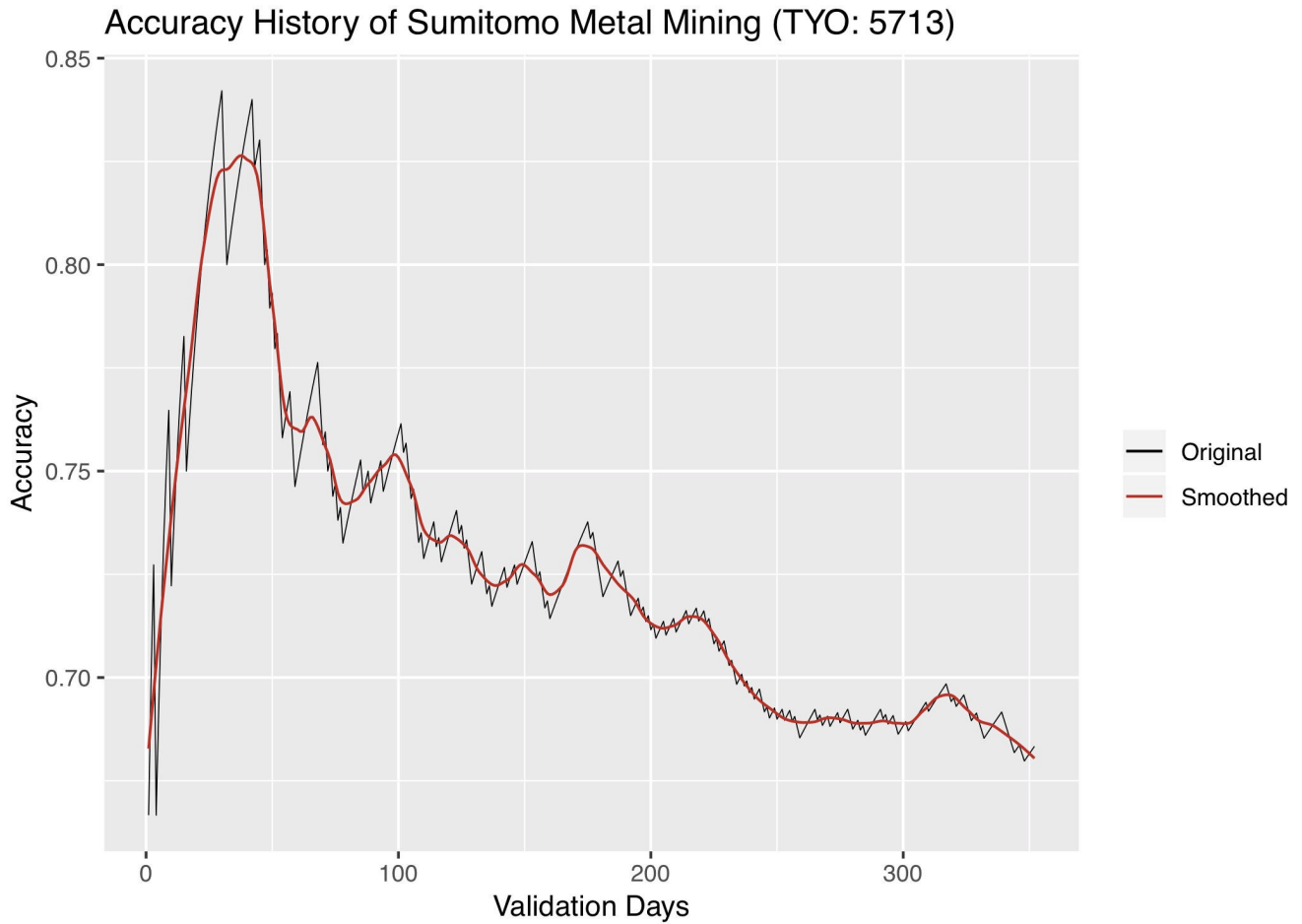

**Fig 13. Accuracy history of Sumitomo.**

https://doi.org/10.1371/journal.pone.0212487.g013

can no longer adapt to pattern changes. When this situation occurs, one should update the training set with newer data and retrain both the base and ensemble models of dynABE.

This observation of accuracy gain and accuracy decay is important for implementing dynABE in practice. As we show in the experiments above, deciding on the right length of validation, or active trading, can effectively improve the accuracy. However, making such decision during training is not intuitive, if not possible. Instead, we suggest choosing the trading length empirically during validation. For example, one can keep track of a live accuracy plot like Figs 12–14, updated every day, and decide to stop trading and retrain the entire model with new data once dynABE starts to show a trend of gradual accuracy decay.

### 5.6 Trading strategy performance

We now use the trend signals that dynABE generates on a naïve trading strategy. Assume that all our assets are cash in the beginning. If the stock price is predicted to rise the next day, we use all our cash to buy shares at the closing price today. If the price is predicted to fall the next day, we sell all the shares we hold at the moment at the closing price today. We record our current asset every day as cash plus the current value of our stock shares. At the end of the trading period, we would obtain a record of asset history using this naïve trading strategy. Then we

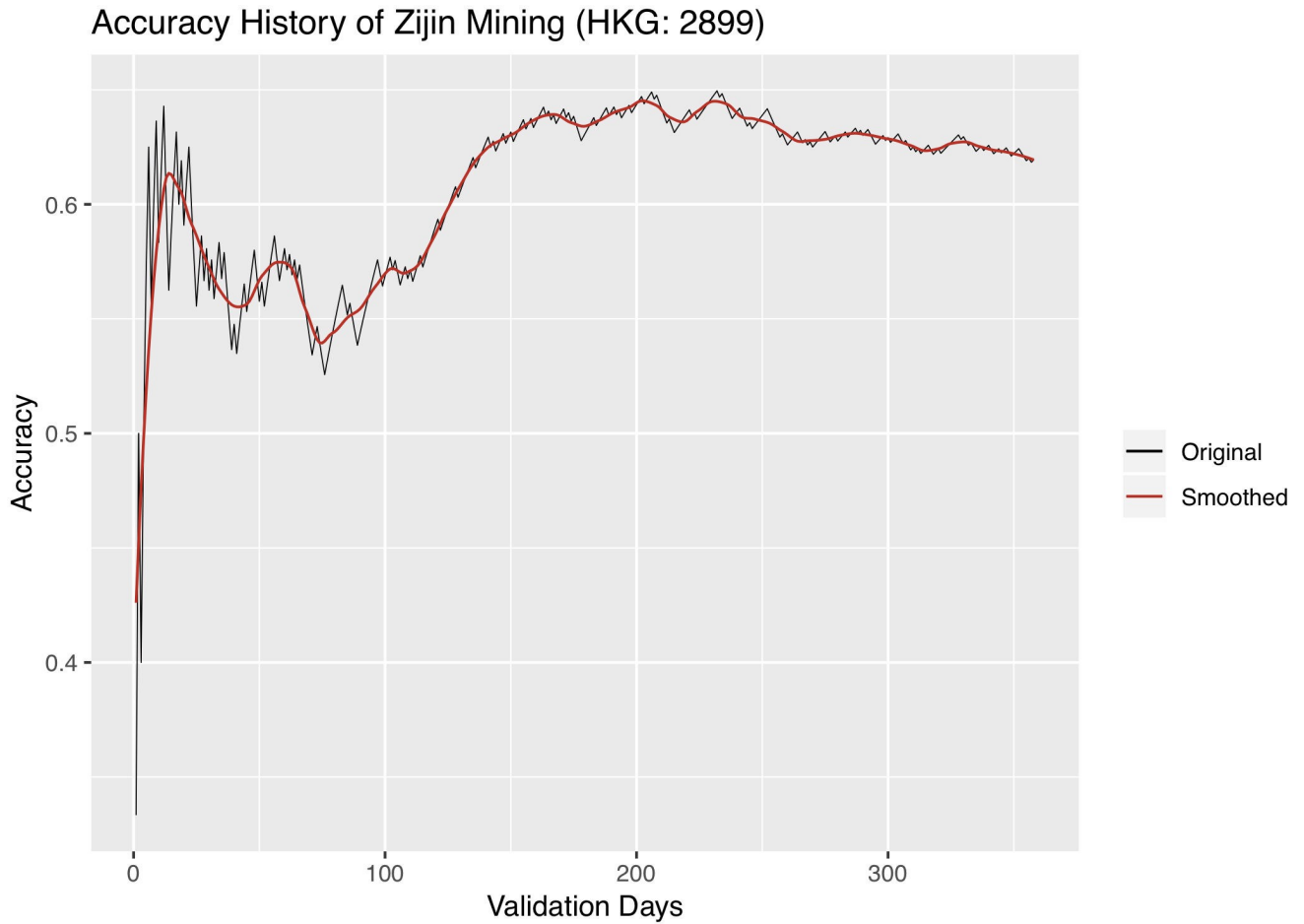

**Fig 14. Accuracy history of Zijin.**

https://doi.org/10.1371/journal.pone.0212487.g014

plot the return rate of our naïve trading strategy with the return rate of the stock itself. Figs 15–17 are the absolute return plots in the trading window of a year and a half. We label the exact percentage of the absolute returns of both the strategy and the stock at the end of each trend line.

Each company has a short weight initialization period in the beginning, represented as a flat black line, and we use the optimal online update predictions for creating these trading strategies.

We evaluate our trading strategy further with financial metrics. Specifically, because our trading period is more than a year, we first annualize the absolute returns using 250 days as the number of trading days in a year. We then show the annualized excess return compared to the stock's own return, which is the difference in the trading strategy's annualized absolute return and the stock's annualized absolute return. Next, we calculate the annualized excess return compared to the stock index of the stock exchange the company's stock is in. Since Jinchuan and Zijin are listed in Hong Kong Stock Exchange, we compare with the Heng Seng Index. And since Sumitomo is listed in Tokyo Stock Exchange, we compare with the JPX-Nikkei 400 Equity ETF Index. In addition, we calculate the Sharpe ratio and the maximum drawdown of the trading strategies. Sharpe ratio is proposed by William Sharpe [49] to measure the expected excess return of a strategy over a risk-free investment per unit of risk [50]. Here we take the US

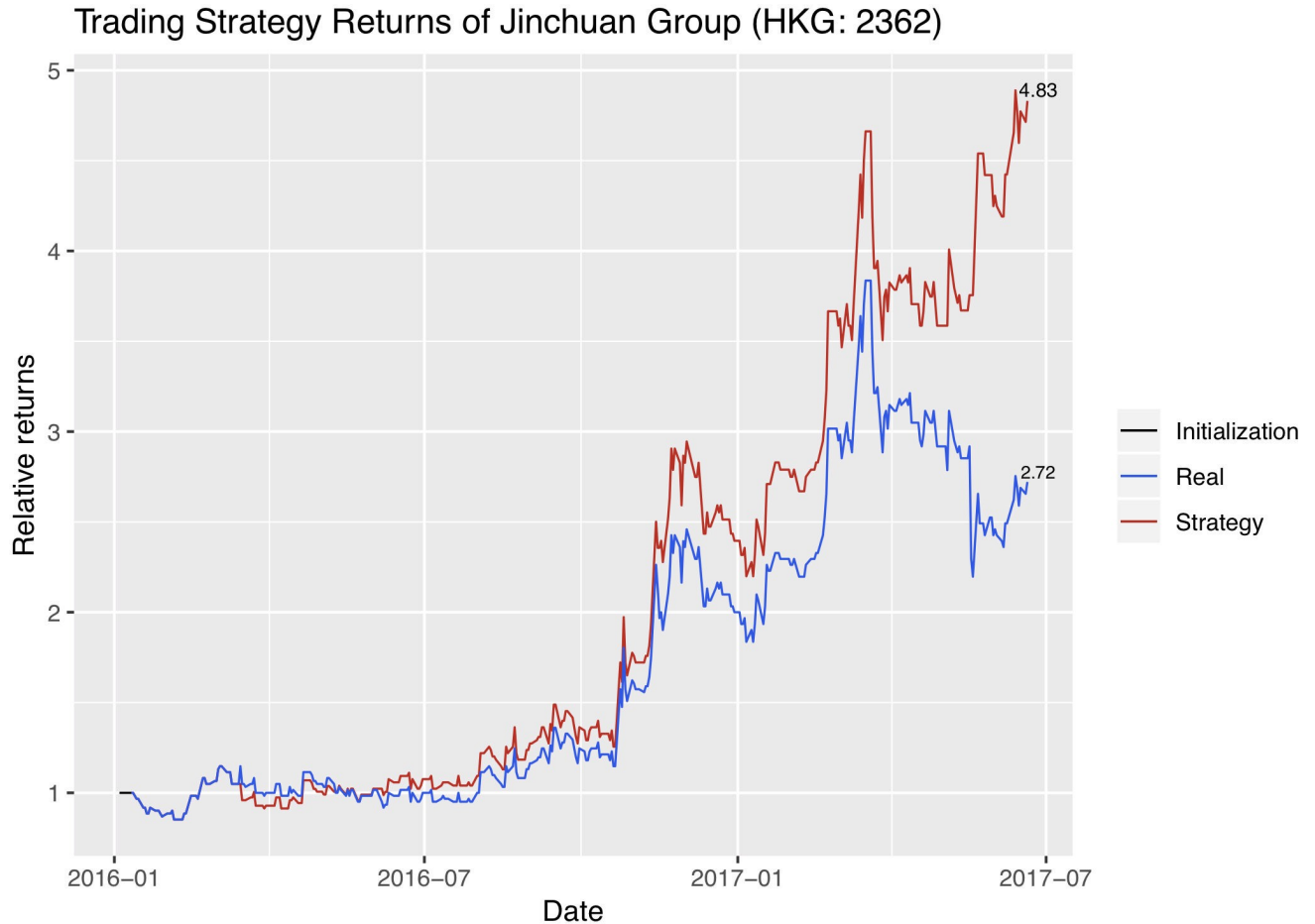

**Fig 15. Absolute returns of the trading strategy and the stock price for Jinchuan.** Exact returns of the last day of the trading period are labled at the end of each trend line. The weight initialization period in the beginning is plotted as a flat black line.

https://doi.org/10.1371/journal.pone.0212487.g015

Treasury's 12-month bond rate to be the risk-free investment. Maximum drawdown is the percentage of maximum loss from peak to trough [51]. Table 8 summarizes the evaluation results on the trading strategies. We see that Sumitomo and Zijin do not have a single loss in the active trading period so their maximum drawdown is 0.

Our trading strategy returns are only for a rough illustration. A more sophisticated trading strategy, such as one which considers multiple stocks simultaneously, can yield higher returns. In addition, we assume that we are always able to trade at every day's closing price, which is not always realistic. Stock prices may further change in practice if we actively enter the market and trade in high volumes. Therefore, one must consider these implications if he or she wants to use the predictions of dynABE as trading signals.

### 5.7 Future work

For future work, we believe that the stacking step for the base models does not yield the most satisfactory results at this point. The stacked predictions can underperform the base models especially when one of the base models is noticeably worse than the others. Therefore, in the future, we would add some sort of pre-stacking base model filter that removes clearly inferior ones before they are stacked. Ideally, we want the stacked predictions to almost always outperform the individual base models.

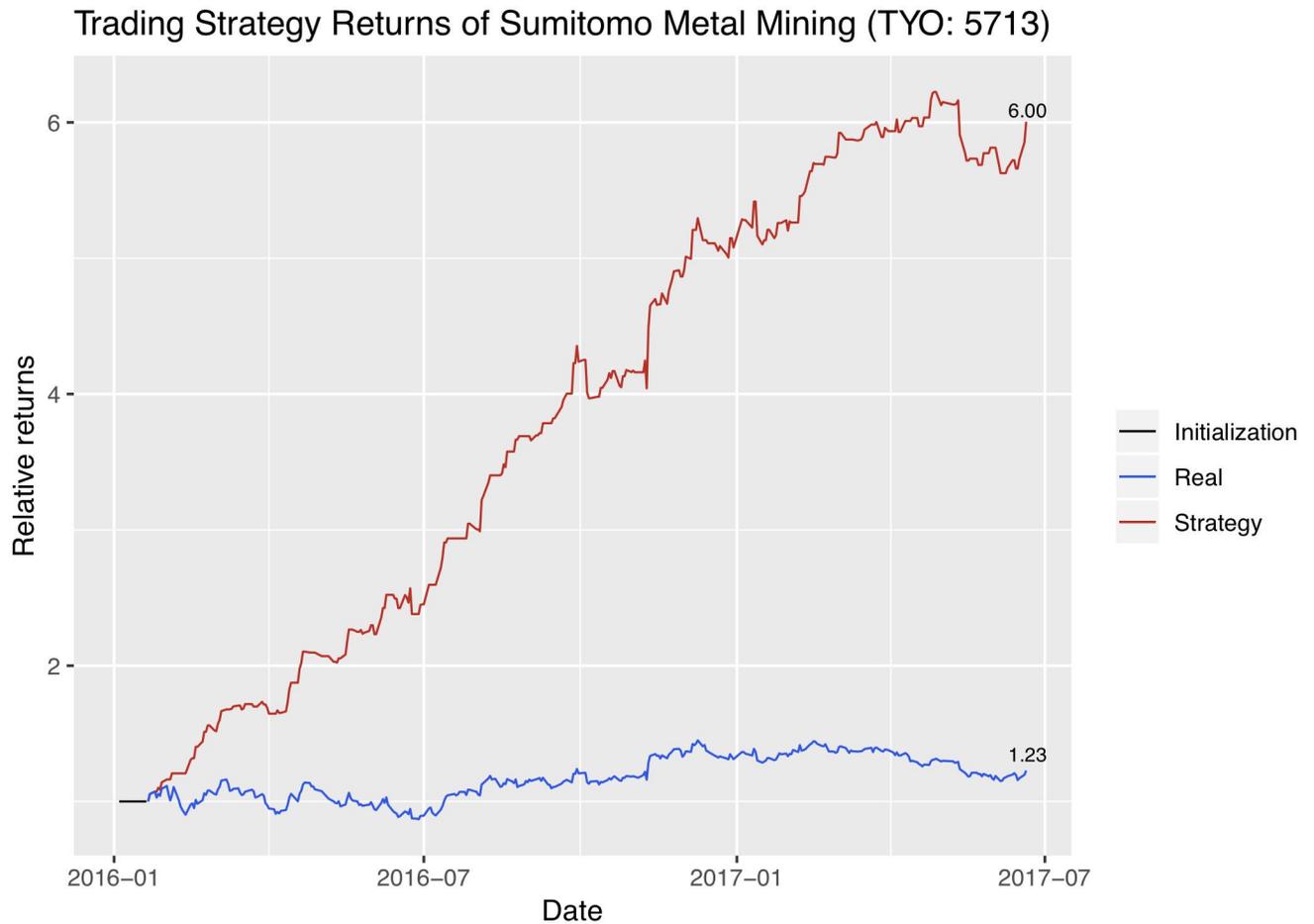

**Fig 16. Absolute returns of the trading strategy and the stock price for Sumitomo.**

https://doi.org/10.1371/journal.pone.0212487.g016

Online update also has a space for improvements. Most importantly, we want to develop a tuning strategy for the update frequency, decay rate, and diversity bias. In addition, while we currently combine the advisors linearly with weighted majority vote, we would like to introduce nonlinearity into combining the advisors in the future. Moreover, similar to a pre-stacking filter, we can also experiment with a pre-online-update filter to filter out clearly poor-performing agents.

## 6 Conclusion

By achieving a best-case misclassification error of 31.12% for Jinchuan Group and a best-case profit of 359.55% annualized absolute return for Sumitomo Metal Mining, dynABE demonstrates accurate stock predictions and high profitability. dynABE also consistently outperforms support vector machine, neural network, and random forest in all case studies. Even though we only investigate critical metal companies in this paper, dynABE can be used for stock prediction of any company. The advantages of dynABE for stock prediction lie in the fact that it uses domain-specific information for advisor creation, relies on an effective ensemble learning framework, and is dynamically adaptive to market changes due to its robust online update strategy.

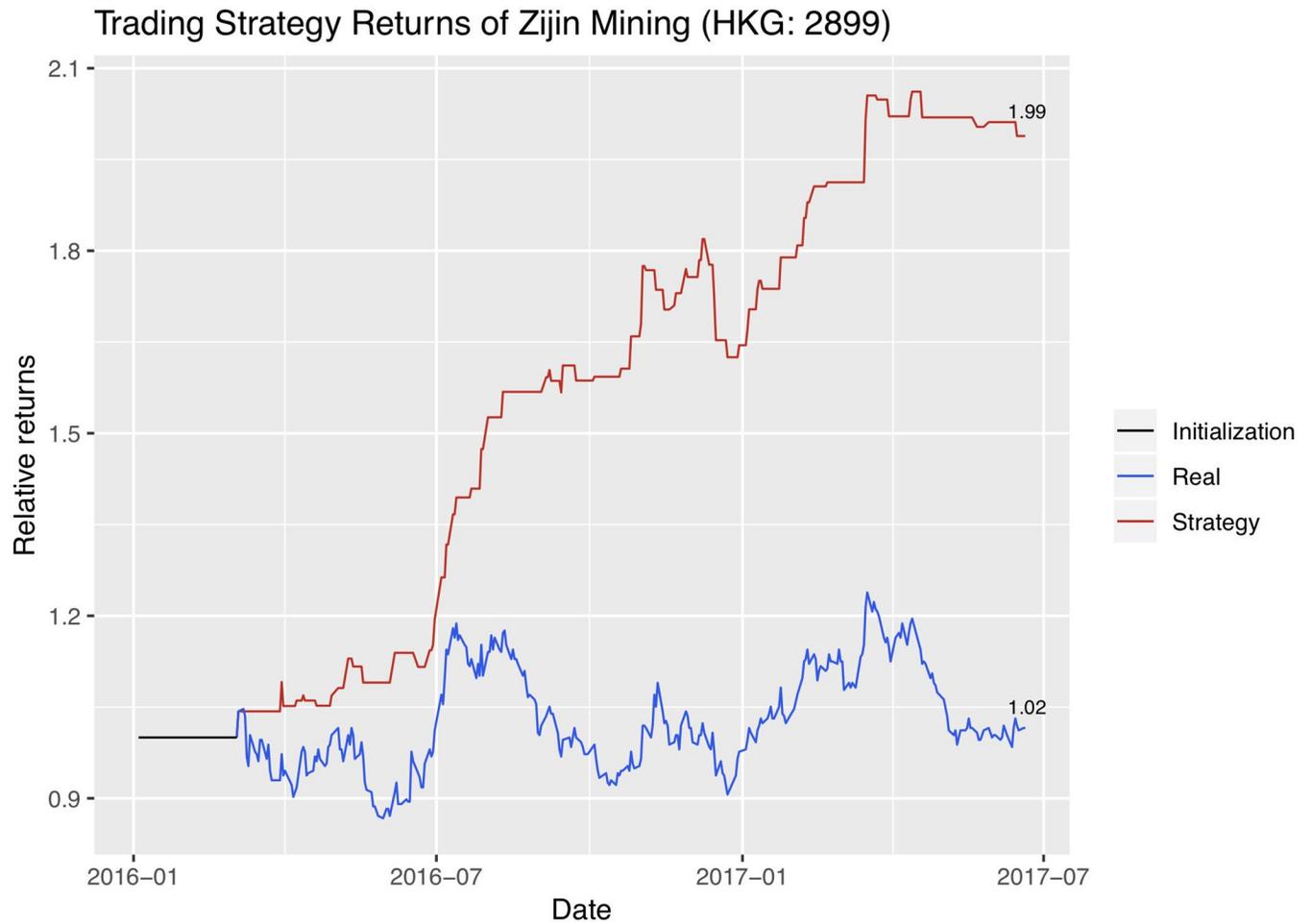

**Fig 17. Absolute returns of the trading strategy and the stock price for Zijin.**

https://doi.org/10.1371/journal.pone.0212487.g017

The various analyses done on the final predictions gain us insights into how dynABE works in practice. Visualization of the online update hyperparameter grid search gives intuitions for developing a tuning strategy in the future. We also plot the weight changes of different advisors and observe weight imbalances in some cases. In addition, we further analyze the accuracy history as the active trading period gets longer and see the interesting phenomenon of accuracy gain and decay, which can help us decide on the optimal trading period before retraining the base and stacking models with new data.

We hope that the ideas of extensively using ensemble learning, creating advisors, and using ideas of decay rate and diversity bias in developing an effective online update strategy can help

**Table 8. Evaluations on trading strategies.** The returns have been annualized using 250 days as the number of trading days in a year.

| Company | Annualized Absolute Return (%) | Annualized Excess Return to Stock (%) | Annualized Excess Return to Index (%) | Sharpe Ratio | Maximum Drawdown |
|---|---|---|---|---|---|
| Jinchuan | 254.704 | 140.255 | 253.840 | 2.08089 | 0.930976 |
| Sumitomo | 359.549 | 343.168 | 358.643 | 2.15309 | 0 |
| Zijin | 77.2329 | 76.0122 | 76.2205 | 2.16598 | 0 |

https://doi.org/10.1371/journal.pone.0212487.t008

future researchers deal more effectively with other similar types of noisy time-series data. Indeed, dynABE is an ensemble learning framework in essence, and it is open to numerous flexible changes to potentially adapt to tasks other than stock prediction.

## Supporting information

**S1 Appendix. Preprocessing procedures.** We provide the details of preprocessing and cleaning the data.
(PDF)

**S1 File. Feature Set descriptions.** This file is a comprehensive table that provides the descriptions of each feature in our original feature set and its symbol.
(CSV)

**S2 File. Symbols of selected features.** This file is a table that shows the features each advisor selects after the feature selection process for all three companies. The features are given in symbols corresponding to descriptions in S1 File. The features are also ranked based on their combined rankings as discussed in the feature selection procedure. The top-ranking feature of each advisor appears first in the table. The suffix of the feature starting with an underscore represents the lag number of this feature. For example, "GSCITOT_1" means that it is the first-lagged version of the feature with symbol "GSCITOT." Each feature was lagged five times in the preprocessing step. See S1 Appendix for details on lagging.
(XLSX)

**S3 File. Stationary analysis of original and first-differenced features.** This file is referenced in S1 Appendix. It contains two tables that show the Augmented Dickey-Fuller (ADF) tests for stationary analysis on the original and first-differenced features respectively. The features of the original feature set that were removed during data cleaning, as described in S1 Appendix, did not run this test.
(XLSX)

## Acknowledgments

The author would like to thank Xinkai Fu from the Department of Materials Science and Engineering, MIT for his assistance on data collection and interpretation. The author would also like to thank Asuka Saito from Nogizaka46 for her motivations when bottlenecks were met during the research process.

## Author Contributions

**Conceptualization:** Zhengyang Dong.

**Formal analysis:** Zhengyang Dong.

**Methodology:** Zhengyang Dong.

**Software:** Zhengyang Dong.

**Validation:** Zhengyang Dong.

**Visualization:** Zhengyang Dong.

**Writing – original draft:** Zhengyang Dong.

**Writing – review & editing:** Zhengyang Dong.

# References

1. Allen H, Taylor MP. Charts, Noise and Fundamentals in the London Foreign Exchange Market. Econ J. 1990; 100: 49. https://doi.org/10.2307/2234183
2. He Z, He L, Wen F. Risk Compensation and Market Returns: The Role of Investor Sentiment in the Stock Market. Emerg Mark Finance Trade. 2019; 55: 704–718. https://doi.org/10.1080/1540496X.2018.1460724
3. Chen Y, Wei Z, Huang X. Incorporating Corporation Relationship via Graph Convolutional Neural Networks for Stock Price Prediction. Proceedings of the 27th ACM International Conference on Information and Knowledge Management—CIKM '18. Torino, Italy: ACM Press; 2018. pp. 1655–1658. https://doi.org/10.1145/3269206.3269269
4. Ou JA, Penman SH. Financial statement analysis and the prediction of stock returns. J Account Econ. 1989; 11: 295–329. https://doi.org/10.1016/0165-4101(89)90017-7
5. Berk RA. An Introduction to Ensemble Methods for Data Analysis. Sociol Methods Res. 2006; 34: 263–295. https://doi.org/10.1177/0049124105283119
6. Fama EF, French KR. Dividend Yields and Expected Stock Returns. J Financ Econ. 1988; 22: 3–25.
7. Pesaran MH, Timmermann A. Forecasting stock returns an examination of stock market trading in the presence of transaction costs. J Forecast. 1994; 13: 335–367. https://doi.org/10.1002/for.3980130402
8. Lee M-C. Using support vector machine with a hybrid feature selection method to the stock trend prediction. Expert Syst Appl. Elsevier Ltd; 2009; 36: 10896–10904. https://doi.org/10.1016/j.eswa.2009.02.038
9. Schumaker RP, Chen H. Textual Analysis of Stock Market Prediction Using Breaking Financial News: The AZFinText System. ACM Trans Inf Syst TOIS. 2009; 27. https://doi.org/10.1145/1462198.1462204
10. Hagenau M, Liebmann M, Neumann D. Automated news reading: Stock price prediction based on financial news using context-capturing features. Decis Support Syst. 2013; 55: 685–697. https://doi.org/10.1016/j.dss.2013.02.006
11. Saad EW, Prokhorov DV, Wunsch DC. Comparative study of stock trend prediction using time delay, recurrent and probabilistic neural networks. IEEE Trans Neural Netw Publ IEEE Neural Netw Counc. 1998; 9: 1456–1470. https://doi.org/10.1109/72.728395 PMID: 18255823
12. Tsang PM, Kwok P, Choy SO, Kwan R, Ng SC, Mak J, et al. Design and implementation of NN5 for Hong Kong stock price forecasting. Eng Appl Artif Intell. 2007; 20: 453–461. https://doi.org/10.1016/j.engappai.2006.10.002
13. Tsai C, Hsiao Y. Combining multiple feature selection methods for stock prediction: Union, intersection, and multi-intersection approaches. Decis Support Syst. Elsevier B.V.; 2010; 50: 258–269. https://doi.org/10.1016/j.dss.2010.08.028
14. Nelson DMQ, Pereira ACM, de Oliveira RA. Stock market's price movement prediction with LSTM neural networks. 2017 International Joint Conference on Neural Networks (IJCNN). Anchorage, AK, USA: IEEE; 2017. pp. 1419–1426. https://doi.org/10.1109/IJCNN.2017.7966019
15. Das S, Behera RK, kumar M, Rath SK. Real-Time Sentiment Analysis of Twitter Streaming data for Stock Prediction. Procedia Comput Sci. 2018; 132: 956–964. https://doi.org/10.1016/j.procs.2018.05.111
16. Ding X, Zhang Y, Liu T, Duan J. Deep Learning for Event-Driven Stock Prediction. Proceedings of the Twenty-Fourth International Joint Conference on Artificial Intelligence (IJCAI 2015). 2015. p. 7.
17. Patel J, Shah S, Thakkar P, Kotecha K. Predicting stock and stock price index movement using Trend Deterministic Data Preparation and machine learning techniques. Expert Syst Appl. Elsevier Ltd; 2015; 42: 259–268. https://doi.org/10.1016/j.eswa.2014.07.040
18. Ballings M, Van Den Poel D, Hespeels N, Gryp R. Evaluating multiple classifiers for stock price direction prediction. Expert Syst Appl. Elsevier Ltd; 2015; 42: 7046–7056. https://doi.org/10.1016/j.eswa.2015.05.013
19. Patel J, Shah S, Thakkar P, Kotecha K. Predicting stock and stock price index movement using Trend Deterministic Data Preparation and machine learning techniques. Expert Syst Appl. 2015; 42: 259–268. https://doi.org/10.1016/j.eswa.2014.07.040
20. Ballings M, Van Den Poel D, Hespeels N, Gryp R. Evaluating multiple classifiers for stock price direction prediction. Expert Syst Appl. 2015; 42: 7046–7056. https://doi.org/10.1016/j.eswa.2015.05.013
21. Investing News Network. Critical Metals Investing News [Internet]. 2018. Available: https://investingnews.com/category/daily/resource-investing/critical-metals-investing/
22. Gunn G, editor. Critical Metals Handbook. American Geophysical Union; John Wiley & Sons, Ltd; 2014. https://doi.org/10.1016/S0065-3233(08)60135-7